\documentclass[journal]{IEEEtran}
\usepackage{graphicx}
\usepackage{cite}
\usepackage{picinpar}
\usepackage{amsmath}
\usepackage{url}
\usepackage{flushend}
\usepackage[utf8]{inputenc}
\usepackage{colortbl}
\usepackage{soul}
\usepackage{multirow}
\usepackage{pifont}
\usepackage{color}
\usepackage{alltt}
\usepackage[hidelinks]{hyperref}
\usepackage{enumerate}
\usepackage{enumitem}
\usepackage{siunitx}
\usepackage{epstopdf}
\usepackage{pbox}
\usepackage{bm}
\usepackage{amssymb}
\usepackage{upgreek}
\usepackage{xcolor}
\usepackage{booktabs}
\usepackage{siunitx}

\begin{document}

\setlength{\topskip}{4pt}
\setlength{\floatsep}{4pt plus 2pt minus 2pt}
\setlength{\textfloatsep}{5pt plus 1pt minus 1pt}
\setlength{\intextsep}{4pt plus 1pt minus 1pt}
\renewcommand{\topfraction}{0.95}
\renewcommand{\textfraction}{0.05}
\renewcommand{\floatpagefraction}{0.9}

\title{Time-delay Control Using a New Nonlinear
Adaptive Law for Cable-Driven Robots}
\author{
	\vskip 1em
	
	Wenbo Gao,
	Yaoyao Wang, Jiawang Chen,
        Wenliang Zhang,
        and Hanzhuo Wang

	\thanks{
		
		Wenbo Gao, Yaoyao Wang and Hanzhuo Wang are with the National Key Laboratory of Science and Technology on Helicopter Transmission, Nanjing University of Aeronautics and Astronautics, Nanjing 210016, China (e-mail: gaowenbo@nuaa.edu.cn; yywang\_cmee@nuaa.edu.cn; nhcmeewhz@nuaa.edu.cn). 

        Jiawang Chen is with the Donghai Laboratory, Zhoushan 316021, China (e-mail:arwang@zju.edu.cn)
        
        Wenliang Zhang is with the China Academy of Machinery Beijing Research Institute of Mechanical and Electrical Technology Co., Ltd., Beijing 100044, China (e-mail: zhangwl@jds.ac.cn).

        Corresponding author: Yaoyao Wang.
        }
}
\maketitle

\begin{abstract}
Cable-driven manipulators exhibit strong nonlinearities and low structural stiffness, which make precise control challenging under time-varying uncertainties and external disturbances. This paper presents a time-delay-estimation (TDE)-based adaptive fractional-order nonsingular terminal sliding mode (AFONTSM) control strategy for cable-driven robots. A robust controller is constructed within a TDE-based model-free framework by combining fractional-order nonsingular terminal sliding mode error dynamics with a fast terminal sliding mode reaching law. The main contribution is a new adaptive law that introduces an adaptive exponential term into the update gain to form a nonlinear adaptive mechanism. This design improves adaptive regulation under different operating conditions by suppressing noise-induced chattering during smooth tracking while preserving or enhancing the adaptive gain during trajectory reversal. Lyapunov analysis proves the ultimate uniform boundedness of the tracking error. Experimental results show that, compared with the baseline method, the proposed controller reduces RMSE by 34.52$\%$ and 31.11$\%$, ITAE by 33.79$\%$ and 32.97$\%$, and ISCT by 6.69$\%$ and 17.77$\%$ for the two joints, respectively. Further comparisons with recently reported adaptive laws demonstrate that the proposed law provides faster adaptive response, more stable gain evolution, and improved chattering suppression. Additional payload tests further verify the robustness and repeatability of the proposed method.
\end{abstract}


\definecolor{blue}{RGB}{30, 90, 200}

\section{Introduction}
% \label{sec:introduction}
\IEEEPARstart{R}{ecently}, owing to their low inertia, high payload-to-weight ratio, and inherent compliance for safe physical human–robot interaction, cable-driven manipulators have attracted considerable attention from both academia and industry in rehabilitation robotics\cite{wu2025adaptive}, aerial manipulation\cite{peng2025dexterous}, and human–robot collaborative systems\cite{huang2020novel}. In contrast to rigid link manipulators, this architecture situates the actuators on a stationary base and transmits power via flexible cables, thereby reducing the arm inertia and improving compliance.

However, the introduction of cable-driven actuation inevitably leads to strong nonlinear behavior and low structural stiffness. During motion reversals, the actuation cables undergo a taut–slack–taut transition due to the unilateral tension constraint and friction-induced hysteresis, which gives rise to piecewise-switching nonlinearities with significant time-varying characteristics. In addition, the manipulator exhibits low stiffness mainly because of cable elasticity and the tension operating point. As a result, during the taut-cable steady-tracking phase, external disturbances are more easily converted into elastic joint deflections, leading to larger tracking errors and high-frequency oscillations. Therefore, the inherent nonlinearities of the system and the time-varying lumped uncertainties and disturbances make high-precision control challenging in practical applications.

To promote the practical application of cable-driven robotic systems, a variety of control strategies have been explored over the past decades. The existing studies have mainly developed model-based controllers within classical servo-control frameworks. Representative examples for cable-displacement tracking include proportional-integral-derivative (PID) control \cite{tran2011control}, model-based open-loop feedforward compensation \cite{sun2015modeling}, and inverse-kinematics-mapping-based PID schemes \cite{tang2019design}. In the joint space, closed-loop regulation has been realized through sensor-based puller-follower proportional control\cite{wu2019closed} and vision-assisted three-loop PD control based on velocity mapping\cite{gu2020experimental}. In the task space, image-based visual servoing has been adopted for end-effector trajectory tracking\cite{wang2013visual}, while tension regulation has also been introduced into cascaded manipulator-control architectures\cite{racioppo2019design}. However, the structural complexity of cable-driven manipulators, together with their inherent hysteresis and disturbance sensitivity, makes accurate modeling highly challenging and places stringent demands on parameter identification.

With the rapid development of intelligent and data-driven control methods, increasing attention has been directed toward the advanced control of continuum manipulators. A fuzzy online tuning baseline computed torque control method was developed by Azita et al.\cite{yazdanpanah2014design} for stable trajectory tracking of continuum manipulators in the presence of parametric uncertainties and external disturbances. For safe motion regulation of a continuum surgical manipulator under external loads and rigid collisions, a shielded multiagent deep reinforcement learning controller was proposed by Ji et al.\cite{ji2021towards}. Wang et al. \cite{wang2021hybrid} proposed a hybrid controller that integrates an offline neural-network-based inverse kinematics model with a position-feedback-based load-state observer to compensate for external-load-induced tracking errors in a continuum surgical robot. Giorelli et al. \cite{giorelli2015neural} proposed a feedforward neural-network-based inverse-statics method for a cable-driven soft arm, improving tracking speed and accuracy.

Despite their potential for performance improvement, the aforementioned methods still face practical limitations. Their high computational complexity imposes stringent real-time requirements, and their effectiveness under uncertainties such as friction and hysteresis depends heavily on parameter tuning and online adaptation. Poor tuning may induce oscillations, overshoot, and tension fluctuations, thereby degrading stability and tracking accuracy while increasing safety and maintenance costs.

In \cite{wang2013robust,kim2017force,baek2017practical}, extensive attention has been devoted to time-delay estimation (TDE) as a promising and effective tool for the compensation of complex model uncertainties and external disturbances in integrated electromechanical systems. Through the deliberate use of delayed inputs and system states, unknown dynamics and lumped uncertainties can be approximated, thereby enabling a model-independent control framework. Owing to its distinctive model-free nature, TDE is well suited to the direct characterization of cable-driven manipulator dynamics, without reliance on computationally intensive and potentially risky optimization procedures. 

Nevertheless, TDE inevitably introduces estimation errors under finite sampling rates, and their practical accuracy is further limited by hardware capability. Accordingly, the integration of TDE with sliding-mode control has become a common remedy for closed-loop stabilization. However, in cable-driven manipulators, the high-frequency switching of sliding-mode control may aggravate chattering, while fixed-parameter laws often struggle to maintain satisfactory performance under time-varying uncertainties. In this context, adaptive control algorithms, characterized by online controller-parameter adjustment in response to environmental variations, have been widely introduced into a variety of integrated electromechanical systems, including cable-driven manipulators\cite{wang2019time}, rigid manipulators\cite{hu2025adaptive,park2023improved} and pneumatic soft actuators\cite{zhao2025adaptive}, due to their potential for rapid response, high-gain regulation, and chattering alleviation. For cable-driven manipulators, Wang et al.\cite{wang2019time} proposed an adaptive control algorithm that improves tracking accuracy and response speed. However, that method still relies on linear updating gains, which limits its ability to balance control performance and chattering suppression under different operating conditions. For rigid manipulators, the adaptive law in \cite{hu2025adaptive} improves the sliding-manifold dynamics and accelerates convergence. However, its update mechanism may still respond insufficiently promptly to variations in the tracking error. Moreover, the adaptive laws reported in\cite{hu2025adaptive,park2023improved,zhao2025adaptive} were all designed to adjust the switching term in the sliding-mode controller. Because the switching term constitutes the discontinuous high-frequency component of sliding-mode control, inadequate adaptation of its coefficient may amplify oscillations and chattering, thereby degrading closed-loop stability.

Motivated by the above analysis, a novel adaptive time-delay control (ATDC) strategy is formulated to enhance the control robustness and tracking precision of cable-driven manipulators.
First, the TDE method is utilized to reconstruct aggregated unmodeled dynamics based solely on measurable system states.
Then, a robust controller is designed within the TDE-based model-free framework by combining FONTSM error dynamics with a fast-TSM reaching law for rapid error attenuation. 
After that, an adaptive update mechanism is formulated to bolster the robustness against time-varying uncertainties and external disturbances. Embedding an adaptive exponential term into the update gain, the proposed adaptive law improves robustness against hard nonlinear disturbances and effectively suppresses the chattering of control signals induced by noise. The new adaptive law improves control performance while simplifying parameter tuning and implementation.
Finally, the ultimate uniform boundedness of the tracking error is rigorously proven through Lyapunov-based analysis. A series of comparative trials are performed to validate the efficacy and superiority of our designed controller.

Building upon existing research, this work contributes to the literature in the following ways:

\begin{enumerate}
% [label=\arabic*) , leftmargin=0pt, itemindent=3em]
    \item  By leveraging the model-free property enabled by TDE, the proposed control scheme requires only a general manipulator dynamic model rather than an accurate model of the cable-driven system, thereby improving generalizability and practical applicability.
    \item  A robust controller is developed by combining FONTSM error dynamics with a fast-TSM reaching law to achieve rapid dynamic response and strong robustness.
    \item A novel adaptive law is proposed by introducing an adaptive exponential term to construct a nonlinear update gain. This design simplifies controller parameter tuning while improving control performance and suppressing noise-induced chattering under different operating conditions.
    \item The ultimate uniform boundedness of the tracking error is established via Lyapunov analysis, and experiments verify the improved steady-state tracking accuracy, adaptive regulation capability, and robustness of the proposed control strategy.
\end{enumerate}

\textit{Remark 1:} The existing adaptive-law designs \cite{wang2019time,hu2025adaptive,park2023improved,zhao2025adaptive} developed for different integrated electromechanical systems have been reviewed above. To address the composite control requirements of cable-driven manipulators, the proposed adaptive law embeds an adaptive exponential term into the update gain. This design improves the overall adaptive performance by enabling a faster response to performance degradation and sufficiently large control parameters when required. Meanwhile, it maintains a more stable switching-term coefficient, which helps suppress control-output chattering induced by disturbances and measurement noise. In addition, the proposed adaptive law simplifies controller-parameter tuning, and its integration with the model-free TDE framework further improves computational efficiency and engineering applicability.

\section{Proposed ATDC For The Manipulator}
\subsection{Brief Review of TDC Scheme}
The formulation for the dynamics of the n-degree-of-freedom cable-driven manipulator is presented below  
\begin{align}
\bm{\tau}_m &= \mathbf{M} \ddot{\mathbf{q}} + \mathbf{C} \dot{\mathbf{q}} 
+ \mathbf{G} + \bm{\tau}_d + \mathbf{J} \ddot{\bm{\phi}} + \mathbf{D}_m \dot{\bm{\phi}}.
\label{eq:dynamics}
\end{align}
where $\mathbf{q},\ \dot{\mathbf{q}},\ \ddot{\mathbf{q}} \in {\Re ^n}$, represent the angle position vector, the angle velocity vector, and the angle acceleration vector of the joints, respectively. $\bm{\phi},\ \dot{\bm{\phi}},\ \ddot{\bm{\phi}} \in {\Re ^n}$ indicate the position vector, the velocity vector, and the acceleration vector of the motor. $\bm{\tau}_m,\ \bm{\tau}_d \in {\Re ^n}$ denotes the torque vectors generated by the motors and the uncertain lumped external disturbances. ${\mathbf{J}},\ \mathbf{D}_m \in {\Re ^n}$ denote the inertia and damping matrices of the motors. ${\mathbf{M}} \in {\Re ^{n \times n}}$ is the symmetric positive definite inertia matrix. $\mathbf{C},\ \mathbf{G} \in {\Re ^n}$ denote the Coriolis matrix and the gravity force vector, respectively. Actually, $\mathbf{M} = \mathbf{M}_0 + \Delta \mathbf{M}$, $\mathbf{C} = \mathbf{C}_0 + \Delta \mathbf{C}$, $\mathbf{G} = \mathbf{G}_0 + \Delta \mathbf{G}$, whereas the first terms $\mathbf{M}_0,\mathbf{C}_0,\mathbf{G}_0$ are the measurable parts and the second terms $\Delta \mathbf{M}, \Delta \mathbf{C},\Delta \mathbf{G}$ represent the uncertainty of the dynamics. Therefore, \eqref{eq:dynamics} can be rewritten as \eqref{eq:dynamics2}
\begin{align}
\bm{\tau}_m &= \mathbf{M}_0\ddot{\mathbf{q}} + \mathbf{C}_0 \dot{\mathbf{q}} + \mathbf{G}_0
+ \mathbf{U} + \bm{\tau}_d + \mathbf{J} \ddot{\bm{\phi}} + \mathbf{D}_m \dot{\bm{\phi}}.
\label{eq:dynamics2}
\end{align} 
where $\mathbf{U}=-\Delta \mathbf{M}\ddot{\mathbf{q}}-\Delta \mathbf{C}\dot{\mathbf{q}}-\Delta \mathbf{G}$ is the time-varying uncertainty, which is bounded by $\|\mathbf{U}\| \le \Delta_U$ \cite{YU20051957}.
\begin{align}
\Delta_U = c_0 + c_1 \|\mathbf{q}\| + c_2 \|\dot{\mathbf{q}}\|^2
\label{eq:delta_F}
\end{align} 
where ${c_0},{c_1},{c_2}$ are the positive constants. By introducing a constant diagonal matrix $\mathbf{\bar{M}} = \operatorname{diag}(\bar{M}_1, \bar{M}_2, \ldots, \bar{M}_n) \in {\Re ^{n \times n}}$, \eqref{eq:dynamics} can be reformulated as \eqref{eq:t_m0}
\begin{align}
\bm{\tau }_m &= \mathbf{\bar M}\ddot{\mathbf{q}} + \mathbf{V}.
\label{eq:t_m0}
\end{align}
where $\mathbf{V} = \bm{\tau}_d + \mathbf{J} \ddot{\bm{\phi}} + \mathbf{D}_m \dot{\bm{\phi}} + (\mathbf{M} - \mathbf{\bar{M}}) \ddot{\mathbf{q}} + \mathbf{C} \dot{\mathbf{q}} + \mathbf{G}$. The value of $\mathbf{V}$ is obtained using the TDE method, as shown below.
\begin{align}
\mathbf{\hat{V}} \triangleq \mathbf{V}(t - l_s) = \bm{\tau}_m(t - l_s) - \mathbf{\bar{M}} \ddot{\mathbf{q}}(t - l_s)
\label{eq:V_hat}
\end{align}
where $l_s$ represents the sampling interval and $\mathbf{\hat{V}}$ serves as an approximation of $\mathbf{V}$.
This work aims to ensure that the system output follows the desired trajectory ${{\bf{q}}_d} \in {\Re ^n}$ with high precision, which means that the tracking error can be represented as ${\bf{e}} = {{\bf{q}}_d} - {\bf{q}}$, ${\bf{e}}\in {\Re ^n}$. Then, the structure of the TDC scheme is formulated as follows: \cite{8186211}
\begin{align}
\bm{\tau}_m \triangleq\ 
&{\mathbf{\bar{M}}(\ddot{\mathbf{q}}_d + \mathbf{K}_d \dot{\mathbf{e}} + \mathbf{K}_p \mathbf{e}}) \nonumber\\
&\quad + { \bm{\tau}_m(t - l_s) - \mathbf{\bar{M}} \ddot{\mathbf{q}}(t - l_s) }.
\label{eq:TDC}
\end{align}
where ${\mathbf{K}}_d = \operatorname{diag}(k_{d1},k_{d2},\ldots,k_{dn})  \in {\Re ^{n \times n}}$ and ${\bm{K}}_p = \operatorname{diag}(k_{p1},k_{p2},\ldots,k_{pn})  \in {\Re ^{n \times n}}$ are positive designed control parameters. The control method does not involve complex system dynamics, thus resulting in a simplified architecture.

Substituting \eqref{eq:V_hat} and \eqref{eq:TDC} into \eqref{eq:t_m0}, the resulting equation is:
\begin{align}
\mathbf{\ddot{e}} + \mathbf{K}_d \mathbf{\dot{e}} + \mathbf{K}_p \mathbf{e} = \bm{\varepsilon}.
\label{eq:error}
\end{align}
where ${\varepsilon _i} =  - \bar M_i^{ - 1}\left( {{{\hat V}_i} - {V_i}} \right)$ denotes the TDE error, which remains bounded if condition \eqref{eq:I} is satisfied, $\mathbf{\bar{M}}$ is properly chosen, and the sampling interval $l_s$ is sufficiently small \cite{5067322}.
\begin{align}
\left\| \mathbf{I} - \mathbf{M}^{-1} \mathbf{\bar{M}} \right\| < 1.
\label{eq:I}
\end{align}

Note that condition \eqref{eq:I} constrains the mismatch between $\mathbf{M}$ and $\mathbf{\bar{M}}$, which is required for the well-posedness of the TDE-based formulation. In a digital implementation, the boundedness of the TDE error further depends on the sampling interval $l_s$, and the variation rate of the estimated term over one sampling period; a sufficiently small $l_s$ makes $\mathbf{V}(t - l_s)$ a valid approximation of $\mathbf{V}(t )$. Meanwhile, the gains $K_p$ and $K_d$ in \eqref{eq:TDC} mainly determine the PD-type tracking error dynamics, whereas overly large gains may increase sensitivity to measurement noise in practice. Under the bounded TDE error, \eqref{eq:error} implies that, with the appropriate choices of $K_p,K_d$, the tracking error e is also bounded.

\textit{\textbf{Lemma 1:}} The fractional integration operators $I_{c + }^\beta $ and $I_{d - }^\beta $ with ${\Re}(\beta) > 0$ are bounded in $L_p(c,d),1 \le p \le \infty$ \cite{kilbas2006theory}
\begin{align}
\| I_{c+}^\beta g \|_p \le N \| g \|_p, 
\| I_{d-}^\beta g \|_p \le N \| g \|_p, 
N = \frac{(d - c)^{\Re(\beta)}}{\Re(\beta)\, |\Gamma(\beta)|}
\label{eq:Lemma1}
\end{align}

\textbf{\textit{Assumption 1:}} The reference trajectory $\mathbf{q}_d$ is assumed to be a bounded, twice continuously differentiable function, with bounded first and second derivatives.

\textbf{\textit{Assumption 2:}} The manipulator operates on a compact set. Hence, the joint position, velocity, and acceleration are bounded, i.e., there exist positive constants $\bar{q}, \bar{v}, \bar{a}$ such that
\begin{equation}
\|\mathbf{q}(t)\| \le \bar{q}, \quad \|\dot{\mathbf{q}}(t)\| \le \bar{v}, \quad \|\ddot{\mathbf{q}}(t)\| \le \bar{a}, \quad \forall t \ge 0.
\label{Assumption 2}
\end{equation}

\textit{Remark 2:} The boundedness argument associated with \eqref{eq:delta_F} follows the structural boundedness assumption in \cite{feng2002non}, where an explicit upper-bounding function is constructed for the lumped uncertainty; further derivation details can be found in \cite{feng2002non}. With Assumption 2 ensuring bounded $\|\mathbf{q}\|$ and $\|\dot{\mathbf{q}}\|$, the right-hand side of \eqref{eq:delta_F} is finite for all $t$, and hence the term in \eqref{eq:delta_F} is bounded.

\textit{Remark 3:} Assumption 1 ensures bounded
$q_d, \dot{q_d}$ and $\ddot{q_d}$, which implies the boundedness of the composite term $\tilde{\varepsilon_i}$ appearing in \eqref{eq:Vdot}. This boundedness is necessary to upper-bound the Lyapunov derivative and to conduct the subsequent case analysis in \eqref{eq:Vdot2}, which finally leads to the ultimate boundedness of the sliding variable $s$. Assumption 2 is introduced in the modeling part to ensure that the upper bound of the lumped uncertainty in \eqref{eq:delta_F} remains finite, which justifies the boundedness premise used by the TDE-based reformulation.

\begin{figure*}[!t]
\centering
\includegraphics[width=\textwidth]{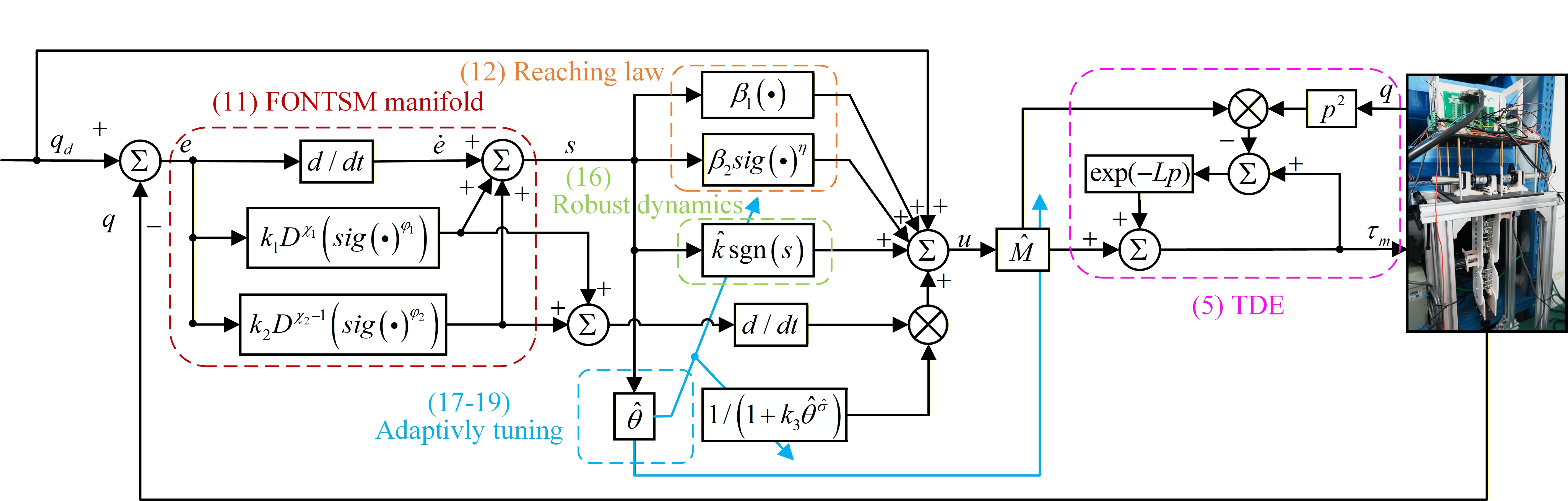}
\caption{Control architecture of the proposed AFONTSMC with TDE.} 
\label{fig:BlockDiagram}
\end{figure*}
\subsection{Structure of the Proposed AFONTSMC with TDE}
The newly proposed robust control approach with TDE utilizes FONTSM error dynamics along with a non-linear control parameter adaptation strategy. The FONTSM manifold and the associated reaching law are formulated to ensure finite-time convergence, as shown below:
\begin{align}
\mathbf{s} &= \dot{\mathbf{e}} + \mathbf{k}_1 D^{\chi_1} \left( \operatorname{sig}(\mathbf{e})^{\varphi_1} \right) + \mathbf{k}_2 D^{\chi_2 - 1} \left( \operatorname{sig}(\mathbf{e})^{\varphi_2} \right) \label{eq:s} \\
\dot{\mathbf{s}} &= - \bm{\upbeta}_1 \mathbf{s} - \bm{\upbeta}_2 \operatorname{sig}(\mathbf{s})^\eta .\label{eq:s_dot}
\end{align}
where $\mathbf{k}_i=\operatorname{diag}(k_{i1},\dots,k_{in}),\bm{\beta}_i=\operatorname{diag}(\beta_{i1},\dots,\beta_{in}),\mathbf{k}_i\in{\Re ^{n \times n}}, \bm{\beta}_i\in{\Re ^{n \times n}},\ k_{in}>0,\ \beta_{in}>0$. The notation $\operatorname{sig}{\left( {\bf{a}} \right)^b} = [|{a_1}{|^{{b_1}}}\operatorname{sign}\left( {{a_1}} \right),\ldots,|{a_n}{|^{{b_n}}}\operatorname{sign}\left( {{a_n}} \right)]$ is used for brevity. $\ 0< \chi_{i1},\varphi_{i1},\dots,\chi_{in},\varphi_{in}<1$ and $\ 0< \eta_{i} <1$ are prespecified control coefficients and the subscript $i$ denotes $i$-th joint $(i=1,2,...,n)$. The operator ${D^{\chi}}[a]$ with $ - 1 < \chi < 1$ represents the fractional calculus operator and the theoretical background is detailed in \cite{LUCHKO2023133906}.

Based on the FONTSM manifold \eqref{eq:s} and a fast-TSM reaching law \eqref{eq:s_dot}, the controller is synthesized as the following structure:
\begin{align}
\bm{\tau}_m &= \mathbf{\hat{M}}\bm{u} + \mathbf{\hat{V}} \nonumber \\
&= \mathbf{\hat{M}}\bm{u} + \bm{\tau}_m(t - l_s) - \mathbf{\hat{M}}\ddot{\mathbf{q}}(t - l_s) \label{eq:t_m4}\\
\bm{u}&= \bm{u}_1 + \bm{u}_2 \label{eq:u} \\
\bm{u}_1&= \ddot{\mathbf{q}}_d + \bm{\xi}(\bm{e}, \bm{\hat{\theta}}) + \bm{\beta}_1 \mathbf{s} + \bm{\beta}_2 \operatorname{sig}(\mathbf{s})^{\bm{\eta}} \label{eq:u1}\\
\bm{u}_2&= \mathbf{\hat{k}}\, \operatorname{sgn}(\mathbf{s})\label{eq:u2}.
\end{align}
where \(
\xi_i = \left( k_1 D^{\chi_{1,i}}(\operatorname{sig}(e_i)^{\varphi_{1,i}}) + k_2 D^{\chi_{2,i}-1}(\operatorname{sig}(e_i)^{\varphi_{2,i}}) \right) \, / 
\)
\\
\(
\left( 1 + k_{3,i} \hat{\theta}_i^{\hat{\sigma}} \right)
\). ${\bf{\hat \theta }}$ is the update gain designed to tune the adaptive parameter ${\bf{\hat M}}$, ${\bf{\hat k}}$.

To enhance robustness against lumped disturbances with time-varying characteristics, an adjustment strategy for the control parameters is introduced, based on adaptive laws driven by the nonlinear update gain $\hat{\theta}_i^{\hat{\sigma}_i}$.
\begin{align}
\hat{M}_i &= \bar{M}_i \left( 1 + k_{3,i} \hat{\theta}_i^{\hat{\sigma}_i} \right), \quad
\hat{k}_i = \bar{k}_i \left( 1 + k_{4,i} \hat{\theta}_i^{\hat{\sigma}_i} \right)\label{eq:M}\\
\hat{\sigma}_i &= \bar{\sigma} \left( 1 - k_{6,i} \hat{\theta}_i \right) \label{eq:sigma}
\end{align}
\begin{equation}
\dot{\hat{\theta}}_i =
\begin{cases}
|s_i|\, \operatorname{sgn}(\hat{\theta}_{i,\text{mid}} - \hat{\theta}_i), & \text{if } \hat{\theta}_i \ge \hat{\theta}_{i,\max} \text{ or } \hat{\theta}_i \le 0 \\
\left( \dfrac{|s_i|}{k_{5,i} \Omega_i} \right)^{\phi_i}, & \text{if } 0 < \hat{\theta}_i < \hat{\theta}_{i,\max},\ |s_i| \ge \Omega_i \\
- (|s_i| + \Delta_i)^{-1}, & \text{if } 0 < \hat{\theta}_i < \hat{\theta}_{i,\max},\ |s_i| < \Omega_i
\end{cases}
\label{eq:thetadot_cases}
\end{equation}
where ${\bar M_i},{\bar k_i},{\bar \sigma _i}$ and ${k_{3-6,i}}$ are pre-specified positive control coefficients of the $i$-th joint. ${\hat \theta _{i,\max }}$ is the upper bound of ${\hat \theta _i}$, while ${\hat \theta _{i,mid}}$ is the intermediate value of ${\hat \theta _{i,\max }}$. ${\Omega _i}$ is a predefined appropriate boundary layer for the sliding manifold. $\phi_i$ and ${\Delta _i}$ are configured to control the change rate of ${\hat \theta _i}$.

\textbf{\textit{Theorem 1:}} Consider the trajectory tracking error and the sliding mode surface ${s_i}$ in \eqref{eq:s}. When $|s_i| \le \bar{\Delta}_i$, the convergence of the trajectory tracking error ${e_i}$ to the region $\Delta_{e_i}$ is ensured by the proposed ATDC scheme. Therefore, the joint angle $q_i$ of the cable-driven manipulators can accurately track the desired trajectory $q_{d,i}$ with the error being ultimately uniformly bounded. The boundedness of ${s_i}$ and $e_i$ is discussed in the Appendix.

The aforementioned theorem shows that the designed controller ensures that the tracking error is ultimately bounded within a small region without highly accurate system models or sophisticated estimation techniques. The control block diagram of the designed AFONTSMC with TDE is illustrated in Fig.~\ref{fig:BlockDiagram}.

\subsection{Discussion on the AFONTSMC with TDE}
\textcolor{blue}{\textit{1) Insights into the AFONTSMC with TDE:}}

An adaptive time-delay control scheme with a model-free structure is formulated for the cable-driven robotic arm system \eqref{eq:dynamics} by incorporating the TDE method, the FONTSM manifold, and a novel adaptive law, as shown in (\ref{eq:s}--\ref{eq:thetadot_cases}). The TDE technique estimates unmodeled system dynamics by exploiting intentionally time-delay state information. The component of injected dynamics, $\bm{u}_1$, employs FONTSM error dynamics and a fast-TSM reaching law. Another component $\bm{u}_2$ is incorporated to improve robustness against aggregated uncertain disturbances. Meanwhile, the control parameters $\mathbf{\hat{M}}$, $\mathbf{\hat{k}}$, $\mathbf{\hat{\sigma}}$ are updated in real time through the newly designed nonlinear adaptive law to enhance control performance. 

\textcolor{blue}{\textit{2) Theoretical Analysis of the Adaptive Law:}}

As we can see from (\ref{eq:M}--\ref{eq:thetadot_cases}), the adaptive gains $\mathbf{\hat{M}}, \mathbf{\hat{k}}, \bm{\hat{\sigma}}$ consist of two parts, the constant term $\bar{M}_i$ and the nonlinear term $\bar{M}_i \cdot k_{3,i} \hat{\theta}_i^{\hat{\sigma}_i}$. The former helps maintain a baseline level of control performance when the lumped disturbances are relatively small and slowly time-varying. The latter plays a more significant role in adaptively providing increased control effort to reinforce robustness and tracking precision when substantial system uncertainties are present. Therefore, the proposed adaptive law yields improved tracking precision and faster transient response compared to the baseline TDE-based scheme. Inspection of (\ref{eq:thetadot_cases}) reveals that $\hat{\theta}_i$ rises as $\dot{\hat{\theta}}_i = (|s_i| / k_{5,i} \Omega_i)^{\phi_i}$ when $0 < \hat{\theta}_i < \hat{\theta}_{i,\max} \text{ and } |s_i| \geq \Omega_i$. $\hat{\theta}_i$ decreases with $\dot{\hat{\theta}}_i = - \left( |s_i| + \Delta_i \right)^{-1}$. The parameter $\Delta_i$ is designed to avoid the singularity issue that may arise when the sliding mode manifold $s_i$ crosses the switching surface. To ensure boundedness, once the update gain $\hat{\theta}_i$ exceeds its predefined limits, it is regulated to decrease or increase in conjunction with the variation of $\dot{\hat{\theta}}_i = \mp |s_i|$.

An adaptive exponential term ${\hat \sigma_i}$ is introduced into the linear update gain ${\hat \theta_i}$, thereby constructing nonlinear term ${\hat \theta _i}^{\hat \sigma_i}$ in the computation of the adaptive control parameters in (\ref{eq:M}). As shown in (\ref{eq:sigma}), the adaptive exponential term ${\hat \sigma_i}$ is derived based on the linear update gain. During the smooth tracking phase, the update gain satisfies $0 \leq \hat{\theta}_{i,\text{smo}} < 1$, remaining relatively small. By pre-designing the parameter $\bar \sigma$ and $k_{6,i}$ such that ${\hat \sigma _{i,\text{smo}}} = \bar \sigma \left( {1 - {k_{6,i}}{{\hat \theta }_{i,\text{smo}}}} \right) > 1$, it naturally results in $0 < \hat{\theta}_{i,\text{smo}}^{\hat{\sigma}_{i,\text{smo}}} \ll \hat{\theta}_{i,\text{smo}}$, which helps to reduce chattering in the update gain caused by noise. During the trajectory reversal stage, the update gain satisfies $0 \leq \hat{\theta}_{i,\text{rev}} < 1$, remaining relatively large. The parameter $\bar \sigma$ and $k_{6,i}$ are designed according to the updated gain ${\hat \theta_i}$ under practical operating conditions, such that $0 \leqslant \hat{\sigma}_{i,\text{rev}} = \bar{\sigma} \left( 1 - k_{6,i} \hat{\theta}_{i,\text{rev}} \right) \leqslant 1$ is satisfied, which consequently ensures that $\hat{\theta}_{i,\text{rev}} \leqslant \hat{\theta}_{i,\text{rev}}^{\hat{\sigma}_{i,\text{rev}}} \leqslant \left[\hat{\theta}_i^{\hat{\sigma}_i}\right]_{\max}$.

The role of the adaptive exponential term can be further illustrated by the following numerical examples:
\begin{enumerate}[label=\arabic*), leftmargin=0pt, itemindent=3em]
	\item Let ${\hat \theta _i} \in \left( {0,0.5} \right)$, $\bar \sigma = 2$, and ${k_{6,i}} = 1$. In the smooth-tracking phase, ${\hat \theta _{i,\text{smo}}} = 0.05$, which gives ${\hat \sigma _{i,\text{smo}}} = 2\left( {1 - 1 \times 0.05} \right) = 1.9 > 1$ and $\hat \theta _{i,\text{smo}}^{{\hat \sigma }_{i,\text{smo}}} = 0.05^{1.9} = 0.00337 \ll {\hat \theta _{i,\text{smo}}}$. This indicates that the nonlinear update gain is significantly reduced, which helps suppress noise-induced chattering. In the trajectory-reversal phase, ${\hat \theta _{i,\text{rev}}} = 0.5$, yielding ${\hat \sigma _{i,\text{rev}}} = 2\left( {1 - 1 \times 0.5} \right) = 1$ and ${\hat \theta ^{{\hat \sigma }_{i,\text{rev}}}}_{i,\text{rev}} = 0.5^1 = 0.5 = {\hat \theta _{i,\text{rev}}}$. Therefore, the adaptive exponential term suppresses chattering in the smooth-tracking phase without weakening the control gain in the trajectory-reversal phase. By contrast, if a constant exponential term is used, then $\hat \theta _{i,\text{smo}}^{{\bar \sigma }} = 0.05^2 = 0.0025$ and ${\hat \theta ^{{\bar \sigma }}}_{i,\text{rev}} = 0.5^2 = 0.25$, which still suppresses chattering in the smooth-tracking phase but undesirably reduces the update gain in the trajectory-reversal phase.
	
	\item Let ${\hat \theta _i} \in \left( {0,0.5} \right)$, $\bar \sigma = 2$, and ${k_{6,i}} = 2$, while keeping all other conditions unchanged. Then, for ${\hat \theta _{i,\text{smo}}} = 0.05$ and ${\hat \theta _{i,\text{rev}}} = 0.5$, the nonlinear update gains are ${\hat \theta ^{{\hat \sigma }_{i,\text{smo}}}}_{i,\text{smo}} \approx 0.0046$ and ${\hat \theta ^{{\hat \sigma }_{i,\text{rev}}}}_{i,\text{rev}} = 1$, respectively. Hence, $\hat \theta _{i,\text{smo}}^{{\hat \sigma }_{i,\text{smo}}} \ll {\hat \theta _{i,\text{smo}}}$ in the smooth-tracking phase, whereas ${\hat \theta ^{{\hat \sigma }_{i,\text{rev}}}}_{i,\text{rev}} > {\hat \theta _{i,\text{rev}}}$ in the trajectory-reversal phase. This result shows that, through appropriate parameter design, the proposed adaptive exponential term can simultaneously suppress chattering during smooth tracking and enhance the adaptive gain during trajectory reversal.
\end{enumerate}

\subsection{Recently Reported Methods}
To facilitate comparison, the structure of a recently proposed scheme\cite{wang2019time} is described below:
\begin{equation}
\begin{aligned}
\bm{\tau}_m =\ & \mathbf{\hat{M}} \Big( \mathbf{\ddot{q}}_d 
+ \boldsymbol{\varpi}(\mathbf{e}, \bm{\hat{\theta}}) 
+ \boldsymbol{\beta}_1 \mathbf{s} 
+ \boldsymbol{\beta}_2\, \operatorname{sig}(\mathbf{s})^{\bm{\eta}} \\
&+\ \mathbf{\hat{k}}\, \operatorname{sgn}(\mathbf{s}) \Big) 
+ \bm{\tau}_m(t - l_s) 
- \mathbf{\hat{M}}(t)\, \mathbf(t - l_s).
\end{aligned}
\label{eq:ATDC_R}
\end{equation}
where ${\varpi_i} = \left( k_{1,i} D^{\chi_{1,i}} \left( \operatorname{sig}(e_i)^{\varphi_{1,i}} \right) \right) / \left( 1 + k_{3,i} \hat{\theta}_i \right) + \left( k_{2,i} D^{\chi_{2,i} - 1} \left( \operatorname{sig}(e_i)^{\varphi_{2,i}} \right) \right) / \left( 1 + k_{3,i} \hat{\theta}_i \right)$. The $s$ is designed as \eqref{eq:s}, and the adaptive law is designed as follows.
\begin{equation}
\hat{M}_i = \bar{M}_i \left( 1 + k_{3,i} \hat{\theta}_i \right), \quad
\hat{k}_i = \bar{k}_i \left( 1 + k_{4,i} \hat{\theta}_i \right)
\label{eq:M_R}
\end{equation}
\begin{equation}
\dot{\hat{\theta}}_i = 
\begin{cases}
\lambda_{1i} |s_i| \operatorname{sgn} ( \theta_{i,\text{mid}} - \hat{\theta}_i ), 
& \text{if } \hat{\theta}_i \geq \hat{\theta}_{i,\max},\ \hat{\theta}_i \leq 0 \\
\lambda_{2i} \left( \frac{|s_i|}{k_{5,i} \Omega_i} \right)^{\phi_i}, 
& \text{if } 0 < \hat{\theta}_i < \hat{\theta}_{i,\max},\ |s_i| \geq \Omega_i \\
- \lambda_{1i} \left( |s_i| + \Delta_i \right)^{-1}, 
& \text{if } 0 < \hat{\theta}_i < \hat{\theta}_{i,\max},\ |s_i| < \Omega_i
\end{cases}
\label{eq:eta_dot_R}
\end{equation}

The recently reported advanced adaptive laws in \cite{hu2025adaptive,park2023improved,zhao2025adaptive} are given by \eqref{eq:theta_update_hu}--\eqref{eq:k_update_hu}, \eqref{eq:k_update_park}, and \eqref{eq:k_update_zhao}, respectively. The adaptive laws reported in \cite{hu2025adaptive} are given by
\begin{equation}
\dot{\hat{\theta}}_i=
\left\{
\begin{array}{ll}
\lambda_{1i}\gamma_i^{|e_i|}\operatorname{sign}(\hat{\theta}_{i,\max}-\hat{\theta}_i), & \text{if }\hat{\theta}_i \le 0 \text{ or } \hat{\theta}_i \ge \hat{\theta}_{i,\max}, \\[2mm]
\lambda_{1i}\gamma_i^{|e_i|}\operatorname{sign}(|e_i|-\Omega_i), & \text{if }0<\hat{\theta}_i<\hat{\theta}_{i,\max}.
\end{array}
\right.
\label{eq:theta_update_hu}
\end{equation}

\begin{equation}
\dot{\hat{k}}_i=
\left\{
\begin{array}{ll}
h_i|s_i|\operatorname{sign}(|s_i|-\omega_i\hat{k}_i^2), & \hat{k}_i>0, \\[2mm]
h_i|s_i|, & \hat{k}_i=0.
\end{array}
\right.
\label{eq:k_update_hu}
\end{equation}

The adaptive law reported in \cite{park2023improved} is given by
\begin{equation}
\dot{\hat{k}}_i=
\left\{
\begin{array}{ll}
h_i|s_i|, & \text{if }|s_i|\ge \Omega_i, \\[2mm]
-l_s^{-1}\hat{k}_i\dfrac{|s_i|}{\Omega_i}, & \text{if }|s_i|<\Omega_i.
\end{array}
\right.
\label{eq:k_update_park}
\end{equation}

The adaptive law reported in \cite{zhao2025adaptive} is given by
\begin{equation}
\dot{\hat{k}}_i=h_i|s_i|^{0.5}\operatorname{sign}(s_i), \quad \text{if } |s_i|\ge \Omega_i.
\label{eq:k_update_zhao}
\end{equation}

The comparative methods considered in this work were selected according to two principles. First, the baseline method in \cite{wang2019time} was chosen because it is the most directly related prior work, developed for cable-driven manipulators within the same TDE-based adaptive control framework. Therefore, it provides the most appropriate controller-level reference for assessing the benefit of the proposed method. Second, the adaptive laws in \cite{hu2025adaptive,park2023improved,zhao2025adaptive} were included because they represent recently reported advanced adaptive update strategies for electromechanical systems and provide meaningful references for evaluating the proposed adaptive mechanism at the adaptive-law level.

A comparative analysis among the baseline method (\ref{eq:ATDC_R}--\ref{eq:eta_dot_R}), the recently reported advanced adaptive laws (\ref{eq:theta_update_hu}--\ref{eq:k_update_zhao}), and the method developed in this work (\ref{eq:t_m4}--\ref{eq:thetadot_cases}) is presented in subsequent sections. At the structural level, a preliminary observation from (\ref{eq:M}), (\ref{eq:sigma}), and (\ref{eq:M_R}) indicates that the proposed adaptive law introduces an adaptive exponential term ${\hat \sigma _i}$, which gives rise to the nonlinear adaptive updating gain ${\hat \theta _i}^{{{\hat \sigma }_i}}$. As can be observed from \eqref{eq:theta_update_hu}, the piecewise formulation contains the term $\gamma_i^{|e_i|}$, whose performance may be sensitive to the order of $|e_i|$. Such dependence may complicate parameter tuning and increase the risk of chattering in the control input. 

The adaptive laws for the switching-term coefficient $\hat k_i$ in (\ref{eq:k_update_hu}--\ref{eq:k_update_zhao}) share a highly similar structure. In all cases, no upper or lower bound is imposed on the integral result of $\dot{\hat{k}}_i$. Consequently, $\hat k_i$ may exhibit persistent high-frequency oscillation, excessive magnitude, or overly rapid variation during the integration process, thereby aggravating chattering in the control system.

Based on the above comparative analysis, the uniqueness and advantages of the proposed adaptive law can be summarized in the following three aspects:

\textcolor{blue}{\textit{1) Reduction of chattering during the smooth tracking phase caused by noise:}} When the adaptive exponential term maintains ${\hat \sigma _i} > 1$ as designed by the adaptive law during the smooth tracking phase, the nonlinear structure ${\hat\theta _i}^{{{\hat\sigma }_i}}$ helps maintain parameter stability and suppress signal chattering.

\textcolor{blue}{\textit{2) Increasing control gain during the trajectory reversal stage to enhance control performance:}} Through proper parameter design, the proposed nonlinear structure is possible to achieve even stronger control performance while ensuring that the control gain remains within the allowable range. 

\textcolor{blue}{\textit{3) Simplifying controller design and reducing deployment complexity:}} By introducing the novel nonlinear structure, the proposed adaptive law is structurally simpler than that in (\ref{eq:eta_dot_R}) and requires three fewer controller parameters. In contrast to the separately designed adaptive laws for the switching-term coefficient in (\ref{eq:k_update_hu}-\ref{eq:k_update_zhao}), the proposed law employs a unified adaptive mechanism for $\hat k_i$, thereby simplifying controller design and reducing tuning effort.

The above statements will be validated through the comparative experiments presented in the following section.

\section{Experimental Verifications}
\subsection{Experimental Platform}
\begin{figure}[!t]
\centering
\includegraphics[width=3.3in]{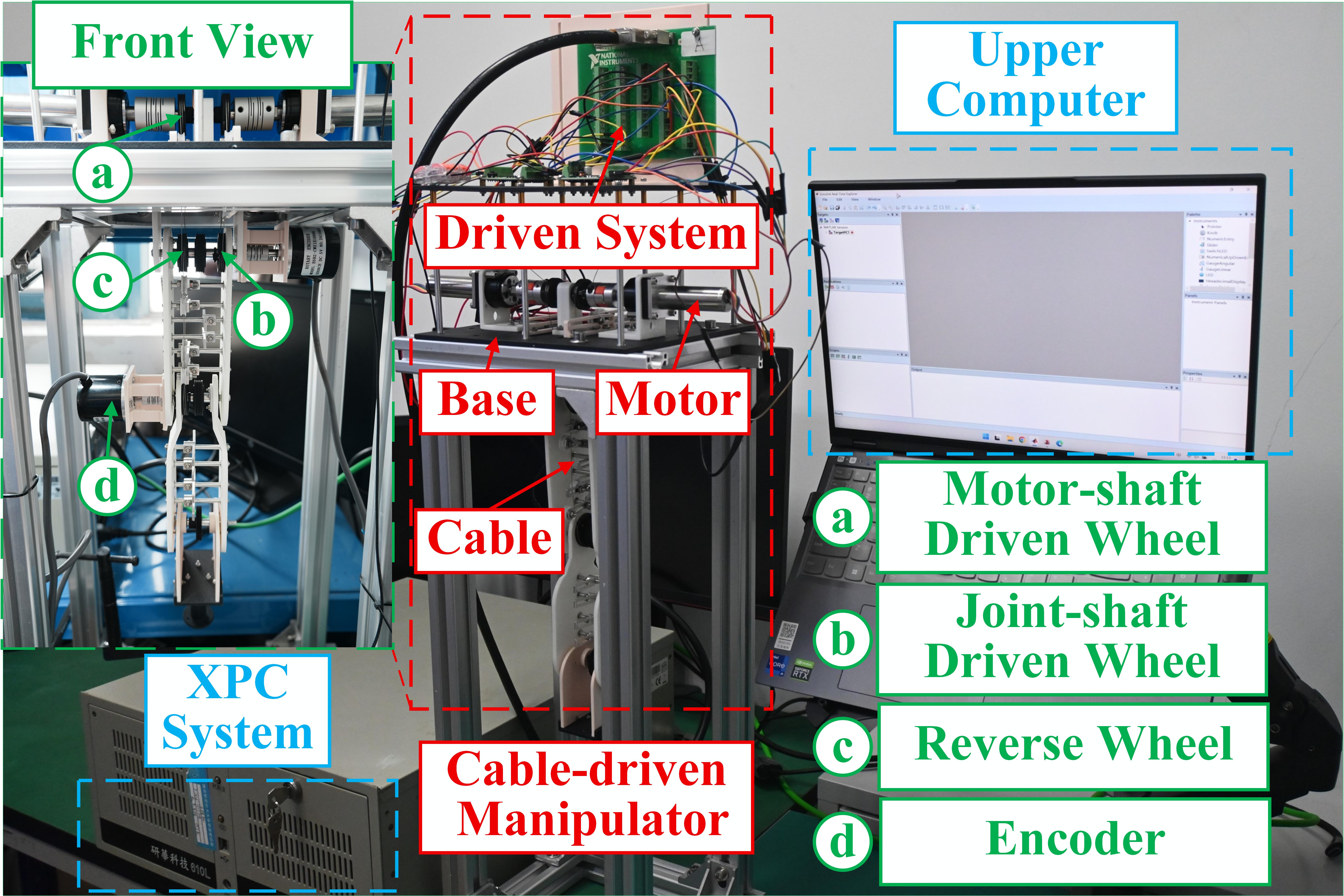}  % 单栏使用宽度 3.3in（约等于 88mm）
\caption{Experimental setup of the cable-driven manipulator.} 
\label{fig:exp_setup_robot}
\end{figure}

Illustrated in Fig.~\ref{fig:exp_setup_robot}, the experimental platform consists of a cable-driven robotic arm actuated by MOONS's ECU19058H24-S001 motors and AQMD2403BLS-M drivers, which deliver a rated torque of 22 $mN \cdot m$ at 12,806 $r/\min $. A SITO CSF-08-2XH-B harmonic reducer with a 100:1 reduction ratio is integrated to enhance motion transmission precision. Position feedback is provided by an E6B2-CWZ1X encoder with 10,000 pulses/rev and a resolution of 0.009$^\circ $. The control scheme is executed in real time via a MATLAB/Simulink Real-Time system interfaced with the NI PCI-6229 IO board, operating at a 1 kHz sampling frequency.

\textit{Remark 4:}
Within the experimental framework, algorithm development, code generation, and compilation were performed on a host PC, whereas deterministic closed-loop execution was carried out on a dedicated real-time target machine. In this study, the host PC was equipped with a 12th Gen Intel® Core™ i9-12900H processor (2.50 GHz) and 16 GB RAM. The real-time target was an Advantech IPC-610-L industrial PC with an Intel® G2120 processor and 2 GB RAM. For replication, a comparable host workstation equipped with a modern multi-core CPU and at least 16 GB RAM, along with a dedicated real-time target capable of sustaining a deterministic 1 kHz loop with the selected I/O hardware, is recommended.

Four categories of experiments were conducted to evaluate the control performance and robustness of the developed ATDC strategy, as well as the regulation capability of the proposed adaptive law. First, the adaptive mechanism was investigated. Second, a comparative experiment was conducted between the proposed algorithm and the baseline method. Third, a comparative study of adaptive-gain adjustment was conducted to evaluate the proposed adaptive law against recently reported advanced adaptive laws. Fourth, robustness was evaluated by attaching an additional 50 g payload to the manipulator.

To ensure fairness, the comparisons were conducted under the same experimental platform, reference trajectory, sampling interval, and evaluation criteria. In Experiment II, the proposed controller and the baseline method in \cite{wang2019time} were implemented under the same parameter-setting principle for the shared control parameters. In Experiment III, the same sliding surface $s_i$ and tracking error $e_i$ obtained in Experiment II were used as common inputs to all adaptive laws. For the proposed method and the baseline method in \cite{wang2019time}, the shared parameter settings were kept the same as those used in Experiment II. For the adaptive laws in \cite{hu2025adaptive,park2023improved,zhao2025adaptive}, only the law-specific parameters were adjusted, while all other settings were kept unchanged. In this way, the comparison focuses on the intrinsic differences in adaptive-law structure rather than on different operating conditions or arbitrary parameter choices.

The following parameter settings were used in implementing the ATDC algorithm:
$\mathbf{\bar M} = 10^{-2} \times \operatorname{diag}(1.5, 1.3)$, $l_s = 2ms$, $\mathbf{k}_{1-2} = \operatorname{diag}(1.5,\ 1.5)$, $\bm{\chi}_1 = 10^{-2} \times \operatorname{diag}(1, 1)$, $\bm{\chi}_2 = \operatorname{diag}(0.99, 0.99)$, $\bm{\varphi}_{1-2} = \operatorname{diag}(0.85, 0.85)$, $\bm{\beta}_1 = \operatorname{diag}(0.2, 0.4)$, $\bm{\beta}_2 = \operatorname{diag}(0.4, 0.4)$, $\bm{\eta}=\operatorname{diag}(0.85,0.85)$, $\mathbf{k}_{3-4} = \operatorname{diag}(1,\ 1)$, $\mathbf{k}_{5-6} = \operatorname{diag}(2,\ 2)$, $\mathbf{\bar k} = 10^{-2} \times \operatorname{diag}(5, 4)$, $\bm{\bar \sigma } = 10^{-2} \times \operatorname{diag}(2, 2)$, $\mathbf{\Omega} = 10^{-2} \times \operatorname{diag}(5, 5)$, $\bm{\phi}=\operatorname{diag}(2,2)$, $\mathbf{\Delta}_1=\operatorname{diag}(2,4)$, $\bm{\hat{\theta}}_{\max} = \operatorname{diag}(0.5, 0.5)$, $\bm{\hat{\theta}}^{\bm{\hat \sigma }}_{\max} = \operatorname{diag}(1, 1)$, $\bm{\hat{\theta}}_{t=0} = \operatorname{diag}(0, 0)$. To avoid an excessively lengthy parameter-by-parameter discussion, only the main tuning guidelines for the key controller gains are summarized below.

\begin{figure*}[!t]             
\centering
\includegraphics[width=7in]{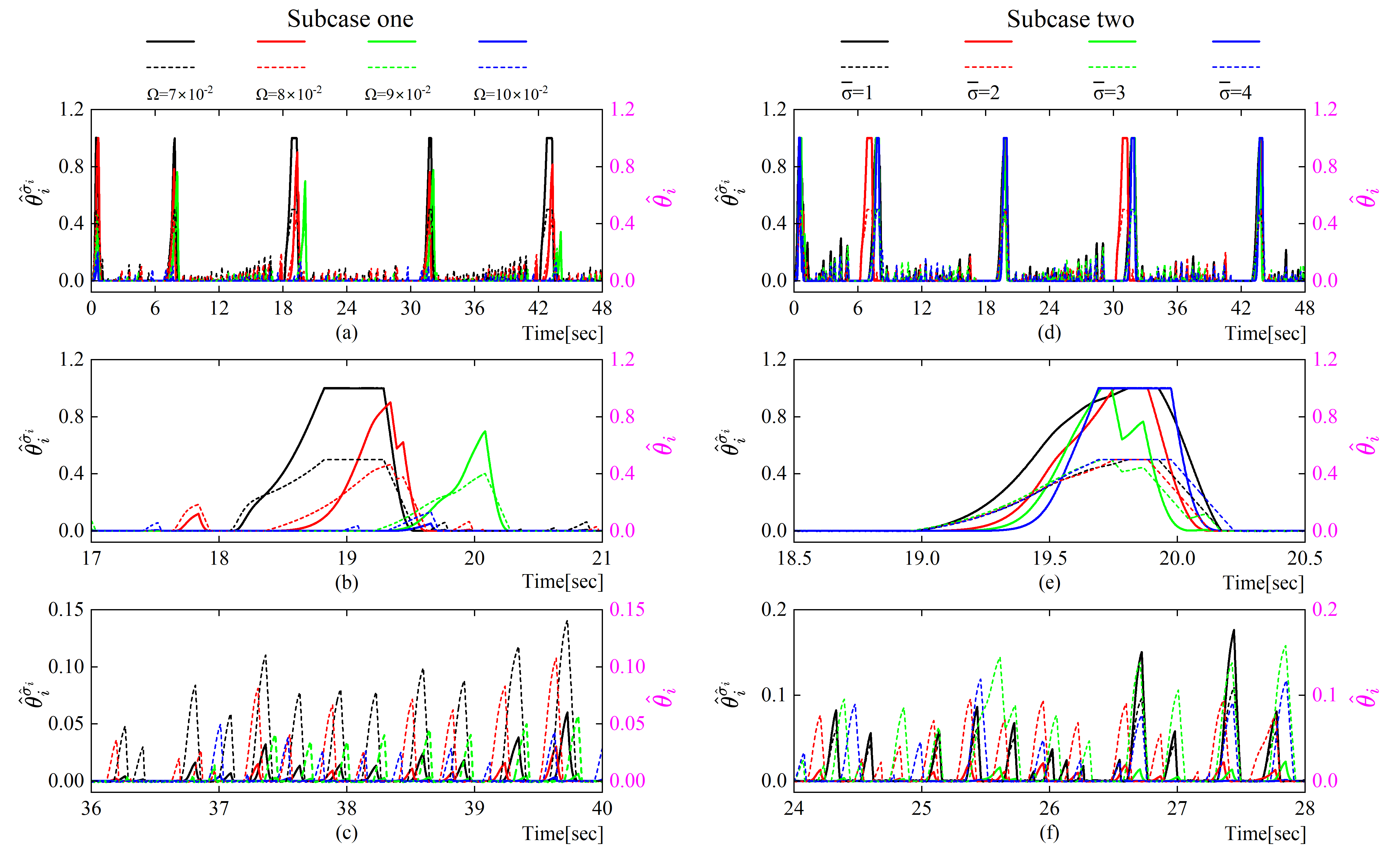}
\caption{Comparison of adaptive nonlinear update gains $\hat{\theta}_i^{\hat{\sigma}_i}$ (solid line) and linear update gains ${\hat \theta _i}$ (dashed line) under the variation of $\Omega$ and $\bar\sigma$. (a) and (d) Entire duration of the experiment; (b) and (e) Zoomed-in view of the trajectory reversal stage; (c) and (f) Zoomed-in view of the smooth tracking phase.}
\label{fig:Comparison of adaptive nonlinear update gains}
\end{figure*}

$\mathbf{\bar M}$ should be chosen such that the TDE mismatch remains bounded and condition \eqref{eq:I} is reasonably satisfied. In practice, its initial value should be set small and then gradually increased until signal chattering appears; the preceding smaller value can be taken as the appropriate setting. The fractional order $\bm{\chi}_{1}$ is chosen as a small positive value so that $D^{\chi_1}(\cdot )$ behaves as a low-order fractional derivative, which retains the memory property of fractional calculus while introducing only a mild differential effect. By contrast, $\bm{\chi}_{2}$ is selected close to unity so that $\chi_2 - 1$ becomes a small negative order, yielding a low-order fractional integral term. The exponents $\bm{\varphi}_{1-2}$ determine the degree of nonlinear error shaping in the manifold. ${\varphi_i}=0.85$ is selected as a compromise between convergence acceleration and chattering suppression: values in $(0,1)$ improve the response to small tracking errors and accelerate convergence near the origin, whereas excessively small values may amplify noise or chattering. $\bm{\beta}_{1}$ mainly affects the linear damping of the sliding variable, whereas $\bm{\beta}_{2}$ and $\bm{\eta}$ determine the nonlinear approaching speed. Increasing $\bm{\beta}_{1-2}$ generally shortens the reaching time, but excessively large values may enlarge the control peaks and aggravate chattering. The boundary layer $\Omega$ defines a practical neighborhood around $s=0$, since the sliding variable cannot be expected to remain exactly at zero in the presence of TDE error, measurement noise, and discrete-time effects. The coefficient $\bm{\bar \sigma }$ together with $\bm{k}_{6}$ determines the adaptive exponential term and governs the trade-off between chattering suppression and adaptive gain preservation. Finally, $\mathbf{\Delta}_1$ avoids singularity in the update law when $s_i$ approaches zero, whereas $\bm{\hat{\theta}}_{\max}$ and $\bm{\hat{\theta}}^{\bm{\hat \sigma }}_{\max}$ are selected according to actuator limitations to prevent excessively large adaptive gains.

\begin{figure*}[!t]                   %3
\vspace*{5pt}
\centering
\includegraphics[width=7in]{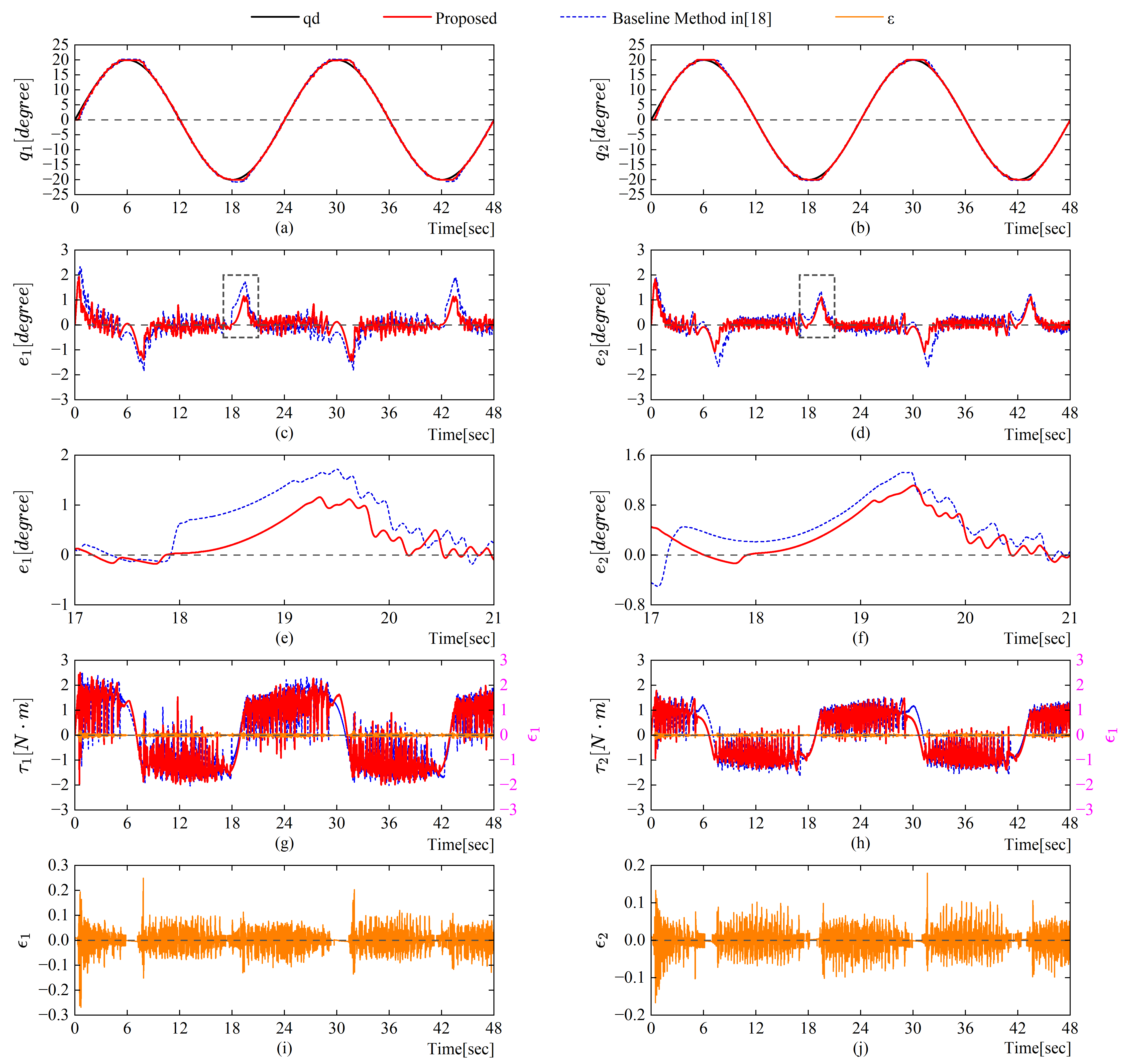}
\caption{Comparison of trajectory tracking performance between the proposed ATDC and the baseline method. (a–b) Trajectory tracking results; (c–d) Tracking errors; (e–f) Zoomed-in views of the trajectory reversal stage; (g–h) Applied control torques of the proposed ATDC and the baseline method, together with the TDE errors ${\varepsilon_i}$ of the proposed ATDC; (i–j) TDE estimation errors ${\varepsilon_i}$ of the proposed ATDC alone.}
\label{fig:Comparison of trajectory tracking performance}
\end{figure*}
\begin{figure*}[!t]                    %4
\centering
\includegraphics[width=7in]{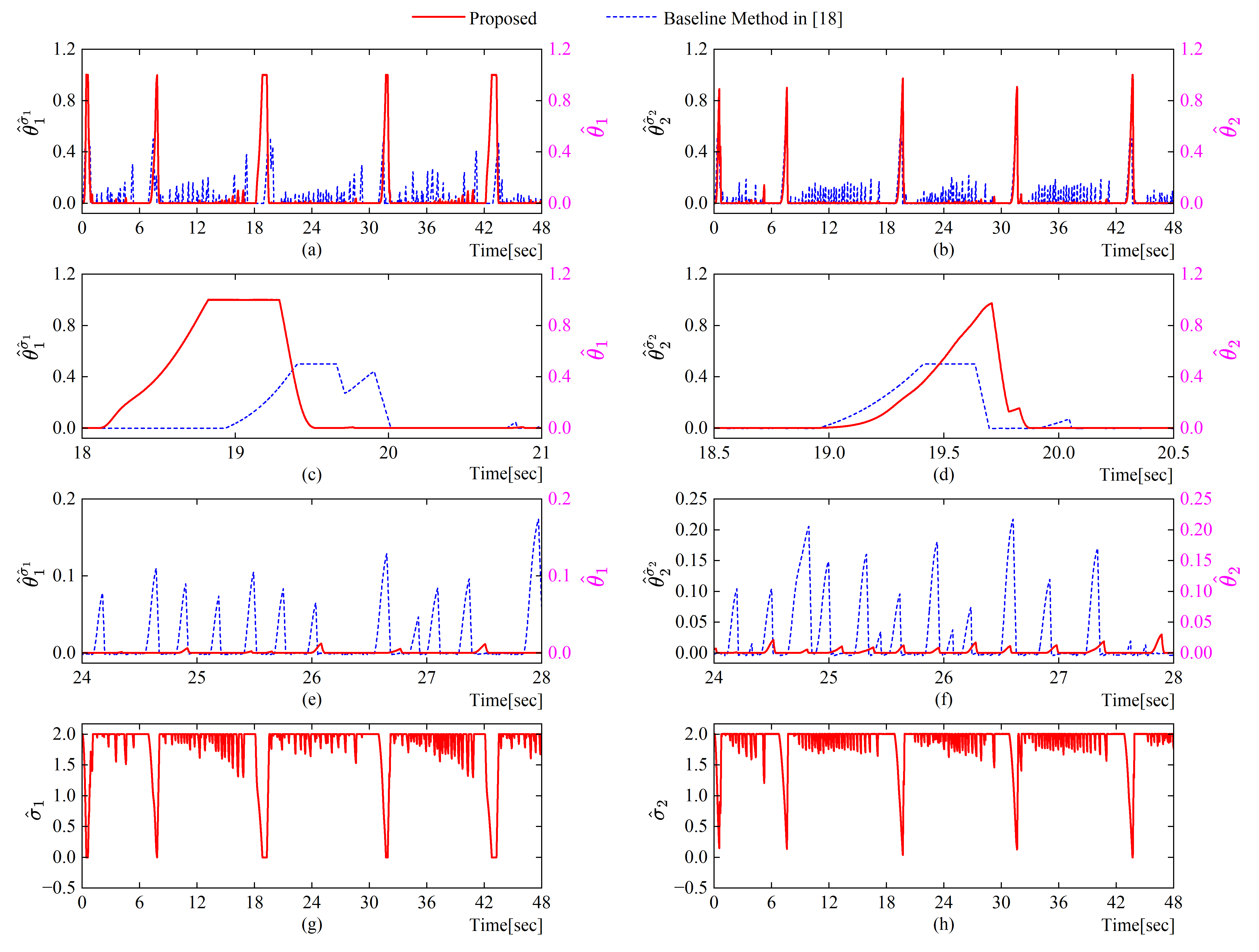}
\caption{Comparison of update gain values between the proposed ATDC ($\hat{\theta}_i^{\hat{\sigma}_i}$) and literature ($\hat{\theta}_i$). (a-b): Entire duration of the experiment; (c-d): Zoomed-in views of the trajectory reversal stage; (e-f): Zoomed-in views of the smooth tracking phase; (g-h): Adaptive exponential term $\hat{\sigma_i}$.} 
\label{fig:Comparison of nonlinear update parameter values}
\end{figure*}

\subsection{Results and Discussion}
\textcolor{blue}{\textit{1) Experiment I - Investigation of the Adaptive Mechanism:}} In this set of experiments, two principle-oriented investigations were conducted. Under the framework of the adaptive exponential term ${\hat\sigma_i}$, the effects of varying boundary layers $\Omega$ and the different exponential coefficient $\bar\sigma$ on the nonlinear update gain ${\hat \theta _i}^{{\hat \sigma }_i}$ were examined. 

\textbf{\textit{Subcase One:}} Mechanism of the boundary layer $\Omega$ variation: The adaptive mechanism was implemented using the parameter settings selected as $\mathbf{k}_{5-6} = \operatorname{diag}(2, 2)$, $\bm{\phi} = \operatorname{diag}(2, 2)$, $\bm{\Delta}_1 = \operatorname{diag}(2, 4)$, $\bm{\hat\theta}_{\max} = \operatorname{diag}(0.5, 0.5)$, $\bm{\hat{\theta}}^{\bm{\hat{\sigma}}}_{\max} = \operatorname{diag}(1, 1)$, ${\bm{\bar\sigma}} = \operatorname{diag}(2,2)$. Four sets of $\mathbf{\Omega}  = {10^{ - 2}} \times \left[ {7,8,9,10} \right]$ were applied in the experiments, with all other controller parameters kept identical to those previously specified. As shown in Fig.~\ref{fig:Comparison of adaptive nonlinear update gains}(a-c), the smaller boundary layer parameter $\Omega$ results in the larger nonlinear adaptive update gain and the longer activation duration. However, it also leads to increased noise sensitivity during the smooth tracking phase. As the boundary layer increases, the control response slows down, with reduced peak values and shorter durations of the $\hat{\theta}_i^{\hat{\sigma}_i}$. Meanwhile, the noise influence during the smooth tracking phase becomes less significant. By comparing the values of the update gain $\hat{\theta}_i^{\hat{\sigma}_i}$ and ${\hat \theta _i}$, it can be observed that the introduction of the adaptive exponential term ${\hat \sigma _i}$ increases the update gain during the trajectory reversal stage while significantly reducing its value during the smooth tracking phase, providing comprehensive control performance.

\textbf{\textit{Subcase Two:}} Mechanism underlying the variation of the exponential term coefficient $\bar \sigma $. Except for the boundary layer parameter $\mathbf{\Omega}  = {10^{ - 2}} \times \left[ {5,5} \right]$ and the four sets of constant exponential terms $\bm{\bar{\sigma}} = [1, 2, 3, 4]$ applied in this subcase, the remaining controller parameters were kept the same as in Subcase One. As shown in Fig.~\ref{fig:Comparison of adaptive nonlinear update gains}(d-f), a smaller value of $\bar \sigma $ provides a faster dynamic response and a longer activation duration. However, when condition $\bar \sigma  = 1$ is met, parameter ${\hat \sigma _i}$ fails to suppress the impact of noise during the smooth tracking phase and instead amplifies the oscillation of the adaptive gain. The reason is that $0 < \hat{\sigma}_{i,\text{fluc}} < 1$ under condition $\bar \sigma  = 1$, which has been analyzed in the preceding discussion. Under condition $\bm{\bar{\sigma}} = [ 2, 3, 4]$, as $\bar \sigma$ increases, the oscillation of the adaptive nonlinear update gain $\hat{\theta}_i^{\hat{\sigma}_i}$ during the smooth tracking phase gradually diminishes, indicating an enhanced ability to suppress noise. 

%%%%%%%%%%%%%%%%%%%%%%%%%%%%%%%%%%%%%%%%

\textcolor{blue}{\textit{2) Experiment II - Validation of the Superior Performance of the Developed ATDC Scheme Over the Existing Baseline Approach:}} For a fair comparison, the proposed controller and the baseline method were evaluated under the same experimental platform, reference trajectory, sampling interval, and evaluation criteria. The shared control parameters were selected according to the same tuning principle. Experimental comparisons illustrating trajectory tracking and control performance are presented in Fig.~\ref{fig:Comparison of trajectory tracking performance}--\ref{fig:Comparison of nonlinear update parameter values}. As shown in Fig.~\ref{fig:Comparison of trajectory tracking performance} and Fig.~\ref{fig:Comparison of nonlinear update parameter values}, both the designed ATDC method and baseline method achieve satisfactory tracking performance with respect to the reference trajectory. By analyzing Fig.~\ref{fig:Comparison of trajectory tracking performance}(c--f), the proposed ATDC method provides improved control accuracy and yields smaller tracking errors. 

To further strengthen the quantitative comparison, four performance metrics are employed: the root mean square error (RMSE), the maximum absolute error (MAE), the integral of time-weighted absolute error (ITAE), and the integral of squared control torque (ISCT). RMSE and MAE are computed over the steady-state interval from $24s$ to $48s$ in order to evaluate the tracking accuracy during periodic steady operation and to reduce the influence of the initial transient stage. ITAE and ISCT are computed over the entire experimental interval from $0s$ to $48s$, because they are used to assess the overall convergence performance and total control effort. Specifically, ITAE penalizes persistent tracking errors, while ISCT quantifies the control effort. For each joint $i$, the ITAE and ISCT are defined as
\begin{equation}
\mathrm{ITAE}_i=\int_0^{T_i} t|e_i|\,dt, \quad
\mathrm{ISCT}_i=\int_0^{T_i} \tau_{m,i}^2\,dt.
\label{eq:ITAE_ISCT}
\end{equation}
where ${T_i}$ denotes the duration of the evaluation interval.

The results are summarized in Table \ref{tab:RMSE_ITAE_ISCT}. For Joint 1, the proposed method reduces the RMSE from $0.5486$ to $0.3592$, corresponding to a $34.52\%$ reduction, while the ITAE is reduced from $415.25$ to $274.92$, i.e., by $33.79\%$. Meanwhile, the ISCT decreases from $76.26$ to $71.16$, corresponding to a $6.69\%$ reduction. For Joint 2, the proposed method reduces the RMSE from $0.4261$ to $0.2935$, corresponding to a $31.11\%$ reduction, while the ITAE decreases from $341.34$ to $228.80$, i.e., by $32.97\%$. The ISCT is also reduced from $40.70$ to $33.47$, corresponding to a $17.77\%$ reduction.

\begin{table}[t]
\centering
\caption{Performance metrics comparison between the proposed method and the baseline method in \cite{wang2019time}.}
\label{tab:RMSE_ITAE_ISCT}
\footnotesize
\setlength{\tabcolsep}{4pt}
\renewcommand{\arraystretch}{1.08}
\begin{tabular}{ccccc}
\toprule
Joint & Case & RMSE & ITAE & ISCT \\
\midrule
   & Proposed          & 0.3592 & 274.92 & 71.16 \\
J1 & Baseline method   & 0.5486 & 415.25 & 76.26 \\
   & Reduction (\%)    & 34.52  & 33.79  & 6.69  \\
\midrule
   & Proposed          & 0.2935 & 228.80 & 33.47 \\
J2 & Baseline method   & 0.4261 & 341.34 & 40.70 \\
   & Reduction (\%)    & 31.11  & 32.97  & 17.77 \\
\bottomrule
\end{tabular}
\end{table}

Overall, the proposed controller yields consistent improvements in RMSE, ITAE, and ISCT for both joints. These results indicate that the proposed approach not only improves tracking accuracy, but also enhances convergence performance while reducing the control effort.

As shown in Fig.~\ref{fig:Comparison of trajectory tracking performance}(g–h), the TDE estimation error ${\varepsilon _i}$  is plotted together with the applied control torque ${\tau _i}$ for both joints. The mismatch remains considerably smaller than the torque over the whole trajectory, indicating that the TDE-based compensation introduces only a limited perturbation to the generated control input. Figs.~\ref{fig:Comparison of trajectory tracking performance}(i-j) depict ${\varepsilon _i}$  alone, where the estimation error stays bounded around zero and exhibits short transient spikes mainly around the start-up and direction-reversal instants. To assess the estimation quality, two normalized indices are introduced:
\begin{equation}
r_{\operatorname{RMS},i}=\frac{\mathrm{RMS}(\varepsilon_i)}{\mathrm{RMS}(\tau_i)}, \quad
r_{\operatorname{MAX},i}=\frac{\max|\varepsilon_i|}{\max|\tau_i|}
\label{eq:error_tao_ratio}
\end{equation}
where $\mathrm{RMS}(\cdot)$ denotes the root mean square value of a signal.

In the experiments, the RMS ratios are ${r_{\operatorname{RMS},1}} = 0.0291$ and ${r_{\operatorname{RMS},2}} = 0.0306$, indicating that the TDE mismatch is small in an energy sense compared with the control effort. The peak ratios are ${r_{\operatorname{MAX} ,1} = 0.1065}$ and ${r_{\operatorname{MAX} ,2} = 0.1004}$, which mainly occur during transient phases. For a more detailed view, we also evaluate the point-wise ratio ${r_i} = |{\varepsilon _i}|/(|{\tau _i}| + \vartheta )$ with a small constant $\vartheta  = 2.2 \times {10^{ - 16}}$ to avoid division by zero. Under a $10\% $ threshold, the exceedance rate is ${r_1} = 7.55\% $ and ${r_2} = 8.18\% $ , and the exceedance is short-lived: the maximum consecutive duration above threshold is $59\,\mathrm{ms}$ and $60\,\mathrm{ms}$ for Joint 1 and Joint 2, respectively (at 1 kHz), with mean durations of about $9-10\,\mathrm{ms}$.

The comparative results in Fig.~\ref{fig:Comparison of nonlinear update parameter values} reveal that the proposed adaptive law exhibits superior overall performance over the baseline method. Specifically, analysis of Fig.~\ref{fig:Comparison of nonlinear update parameter values}(c--f) reveals that our adaptive algorithm produces larger update gain during the trajectory reversal stage. Meanwhile, during the smooth tracking phase, the proposed method effectively suppresses chattering in the control signal caused by measurement noise. Additionally, in Fig.~\ref{fig:Comparison of nonlinear update parameter values}(c--d), the update gain remains constant for a period during the trajectory reversal stage. This is because the parameter reaches the predefined upper bound, which avoids excessively large control signals that may compromise control performance.

\textcolor{blue}{\textit{3) Validation of the Superiority of the Proposed Adaptive Law over Existing Schemes:}} The results of Experiment II have already verified finite-time convergence of the tracking error and stable trajectory-tracking performance under the proposed controller. To further substantiate the core novelty of the proposed adaptive law and ensure a fair comparison, the sliding surface $s_i$ and the corresponding tracking error $e_i$, both obtained from the same run of Experiment II, are employed as the inputs to the proposed adaptive law and several recently reported advanced adaptive schemes \cite{wang2019time,hu2025adaptive,park2023improved,zhao2025adaptive} according to their respective formulations. A comparative investigation is then conducted through graphical analysis of the evolution profiles of parameters $\hat{\theta}_i^{\hat{\sigma}_i}, \hat{\theta}_i$ and $\hat k_i$. For the sake of a fair comparison, the parameter settings of the proposed algorithm and the method in \cite{wang2019time} were kept identical to those used in Experiment II. For the adaptive laws in \cite{hu2025adaptive,park2023improved,zhao2025adaptive}, only a few law-specific parameters were separately tuned as $\mathbf{\gamma}=\operatorname{diag}(0.1,\,0.2)$, ${\bf{h}} = {\operatorname{diag}}(1,1)$, and ${\mathbf{\omega }} = {\operatorname{diag}}(15,20)$, while all the remaining parameters were kept the same as those used in Experiment II.

\begin{figure}[!t]                  
\centering
\includegraphics[width=3.5in]{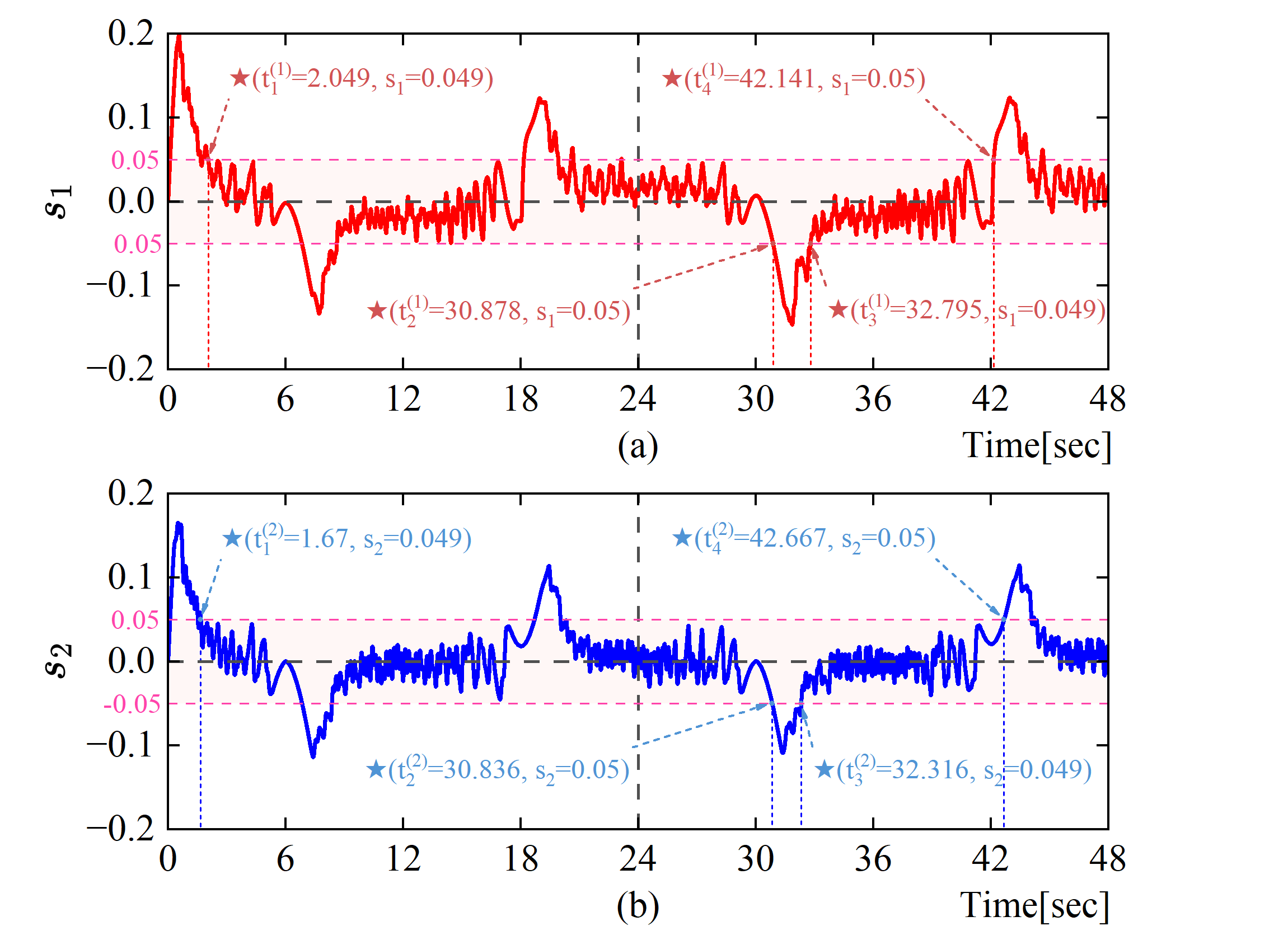}  % 单栏图推荐宽度为 3.3in
\caption{Sliding-surface profiles obtained in Experiment II. (a) Sliding surface $s_1$ of Joint 1 (red solid line); (b) sliding surface $s_2$ of Joint 2 (blue solid line). The dashed magenta lines denote the boundary layer $\Omega_i = 0.05$.}
\label{fig:Sliding surface}
\end{figure}

First, the sliding-surface profile obtained in Experiment II is presented in Fig. \ref{fig:Sliding surface}, where the dashed magenta lines indicate the boundary layer ${\Omega _i} = 0.05$. During the reaching phase, both sliding surfaces rapidly decrease and enter the boundary layer at $t_1^{(1)} = 2.049s$ (Joint 1) and $t_1^{(2)} = 1.670s$ (Joint 2), demonstrating convergence to a neighborhood of zero. After periodic steady operation is established (marked at $t = 24s$), short excursions outside the boundary layer are observed mainly around direction reversals. As highlighted in Fig. \ref{fig:Sliding surface}, Joint 1 leaves the boundary layer at $t_2^{(1)} = 30.878s$ and re-enters $t_3^{(1)} = 32.795s$, yielding a recovery time $T_{rec}^{(1)} = t_3^{(1)} - t_2^{(1)} = 1.917s$; it then stays within the boundary layer until $t_4^{(1)} = 42.141s$, corresponding to $T_{in}^{(1)} = t_4^{(1)} - t_3^{(1)} = 9.346s$. Similarly, Joint 2 leaves the boundary layer at $t_2^{(2)} = 30.836s$ and re-enters at $t_3^{(2)} = 32.316s$, giving $T_{rec}^{(2)} = 1.480s$, and remains bounded within the boundary layer until $t_4^{(2)} = 42.667s$, i.e., $T_{in}^{(2)} = 10.351s$. Overall, the results verify that $s_i$ converges to and stays bounded in a small neighborhood of zero, consistent with the intended sliding-mode behavior and the stability analysis.

\begin{figure*}[!t]                
\centering
\includegraphics[width=7in]{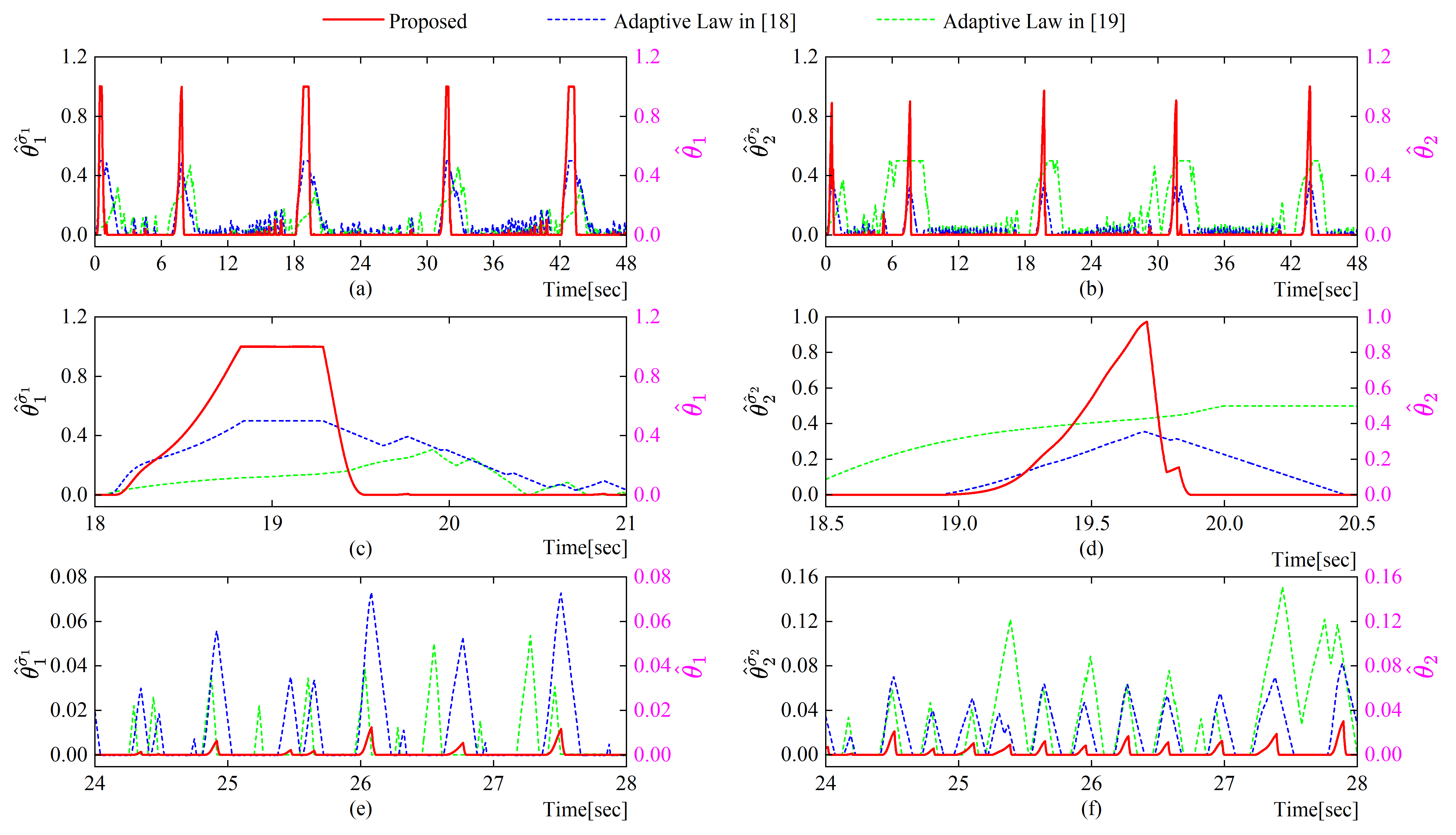}
\caption{Comparison of update-gain profiles under the proposed adaptive law ($\hat{\theta}_i^{\hat{\sigma}_i}$) and two existing adaptive laws ($\hat{\theta}_i$), \cite{wang2019time,hu2025adaptive}. (a–b): Entire duration of the experiment; (c–d): Zoomed-in views of the trajectory-reversal stage; (e–f): Zoomed-in views of the smooth-tracking phase.}
\label{fig:update_gain_adaptive}
\end{figure*}

\begin{figure*}[!t]                
\centering
\includegraphics[width=7in]{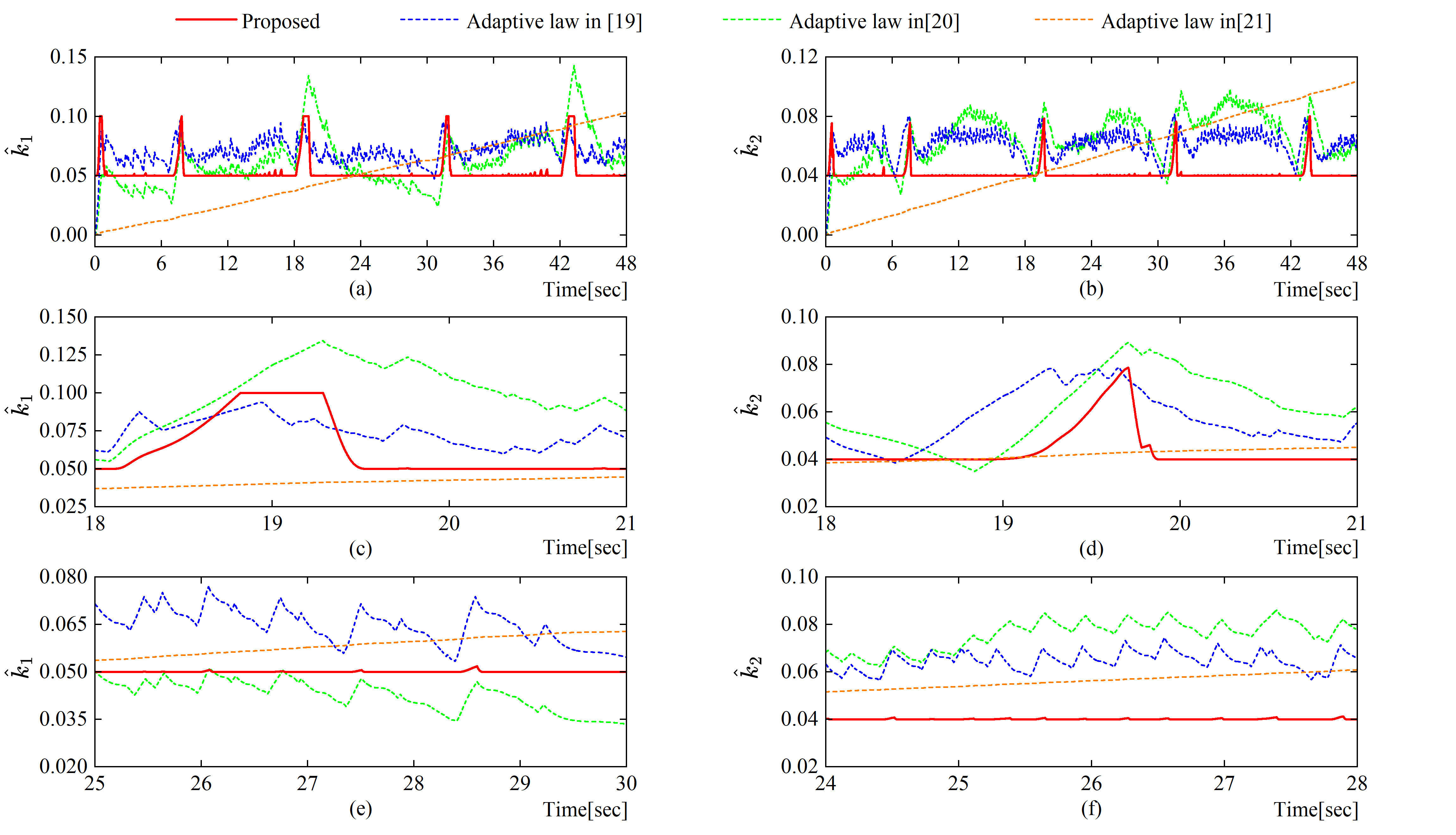}
\caption{Comparison of adaptive-parameter $\hat{k}_i$ profiles under the proposed adaptive law and three existing adaptive laws \cite{hu2025adaptive,park2023improved,zhao2025adaptive}. (a–b): Entire duration of the experiment; (c–d): Zoomed-in views of the trajectory-reversal stage; (e–f): Zoomed-in views of the smooth-tracking phase.}
\label{fig:k_hat_adaptive}
\end{figure*}

The evolutions of $\hat{\theta}_i^{\hat{\sigma}_i}$ and $\hat{\theta}_i$ under different piecewise adaptive laws are illustrated in Fig. \ref{fig:update_gain_adaptive}. The enlarged views in Fig. \ref{fig:update_gain_adaptive}(c--f) further reveal the detailed behaviors in different operating phases. It can be observed that the proposed adaptive law exhibits a faster parameter response and higher gain-adjustment efficiency during the motion-reversal stage. The gain fluctuation during the smooth tracking stage is the smallest among the compared methods. These observations imply improved smoothness of parameter evolution and indicate a greater potential for suppressing control-signal chattering.

As shown in Fig. \ref{fig:k_hat_adaptive}, different evolution patterns of ${\hat k_i}$ are produced by different adaptive laws. For the compared methods, the adaptive parameters are generally initialized from values close to zero, and no explicit upper or lower bounds are imposed on the integral results. As a consequence, persistent oscillation, gradual drift, and even inappropriate amplitude enlargement can be observed during parameter updating. Such phenomena are evident in the global responses in Fig. \ref{fig:k_hat_adaptive}(a)–(b) and are further highlighted by the enlarged views in Fig. \ref{fig:k_hat_adaptive}(c)–(f). In particular, during the smooth tracking stage, laws in \cite{hu2025adaptive} and \cite{park2023improved} still exhibit pronounced high-frequency fluctuations, whereas law in \cite{zhao2025adaptive} shows a monotonic growth trend caused by cumulative integration. These undesirable parameter variations may further amplify the switching-term magnitude, thereby aggravating chattering in the control input.

By contrast, the proposed adaptive law exhibits a clear on-demand regulation mechanism. During the motion-reversal stage, ${\hat k_i}$ are rapidly elevated to larger values, thereby strengthening adaptation to abrupt operating-condition variations. During the smooth tracking stage, however, both parameters quickly return to low and nearly constant levels. This behavior indicates that the proposed method is capable of increasing the adaptive gain when external conditions change, while suppressing unnecessary parameter oscillation as much as possible in smooth tracking  phases. As a result, a better balance between fast adaptability and chattering alleviation is achieved.

\textcolor{blue}{\textit{4) Experiment IV - Validation of Algorithm Robustness:}} Robustness was further tested by attaching a 50 g payload to the end-effector, introducing an additional disturbance for the ATDC algorithm. To further support the robustness of the proposed controller under payload conditions, the payload experiment was repeated three times, and the tracking errors were statistically summarized in terms of RMSE and MAE for both joints. The repeated-run results are reported in Table \ref{tab:payload_repeatability}. The relatively small standard deviations and coefficients of variation indicate that the proposed method exhibits consistent tracking performance across repeated payload trials, thereby supporting the robustness claim.

\begin{table}[t]
\centering
\caption{Repeated-run tracking performance under the \SI{50}{\gram} payload condition.}
\label{tab:payload_repeatability}
\footnotesize
\setlength{\tabcolsep}{3pt}
\renewcommand{\arraystretch}{1.06}
\begin{tabular}{cccccc}
\toprule
Joint & Metric & Run 1 & Run 2 & Run 3 & Mean $\pm$ Std (CV\%) \\
\midrule
J1 & RMSE & 0.4193 & 0.4216 & 0.4046 & 0.4152 $\pm$ 0.0092 (2.22) \\
J1 & MAE  & 1.5354 & 1.6516 & 1.5163 & 1.5678 $\pm$ 0.0732 (4.67) \\
J2 & RMSE & 0.3437 & 0.3238 & 0.3381 & 0.3352 $\pm$ 0.0103 (3.06) \\
J2 & MAE  & 1.3863 & 1.3634 & 1.3931 & 1.3809 $\pm$ 0.0156 (1.13) \\
\bottomrule
\end{tabular}
\end{table}

In summary, the proposed ATDC method and adaptive law achieve superior control performance compared to the existing approaches. The high tracking accuracy, effective adaptivity, and strong robustness have been clearly demonstrated through the mechanism analysis experiments and comparative studies presented above.

\subsection{Limitations and Practical Considerations}

\textcolor{blue}{\textit{1) Computational complexity:}} The proposed controller is developed within the TDE framework and does not require an accurate rigid-body dynamics model or online parameter identification, which reduces modeling effort and implementation burden. The main runtime cost arises from the numerical realization of the fractional-order operators. With a short-memory approximation of length $N_h$, the per-step computational cost and memory requirement scale as $O(nN_h)$. In the present experiments, the controller was successfully executed in real time at 1 kHz on the MATLAB/Simulink Real-Time platform.

\textcolor{blue}{\textit{2) Measurement noise and filtering:}} Due to the low stiffness of cable-driven manipulators, encoder-based joint position measurements ${\bf{q}}$ are susceptible to high-frequency noise. Since the proposed controller involves numerical differentiation to obtain ${\bf{\dot q}}$, mild low-pass filtering is generally required in practice. However, overly aggressive filtering may introduce phase lag and degrade transient tracking performance.

\textcolor{blue}{\textit{3) Sampling rate requirement:}} The proposed scheme relies on the TDE approximation ${\bf{\hat V}} = {\bf{V}}(t - {l_s})$, whose accuracy depends on a sufficiently small sampling interval $l_s$. If the sampling rate is too low or the disturbance varies too rapidly, the TDE error may increase, leading to degraded tracking performance. Therefore, practical implementation requires a proper trade-off among sampling rate, filter bandwidth, and control gains.

\section{Conclusion}
This paper presented an ATDC strategy for cable-driven manipulators. By combining a TDE-based model-free framework, FONTSM error dynamics, a fast terminal sliding mode reaching law, and a new adaptive law with an adaptive exponential term, the proposed method improves adaptive regulation under different operating conditions and enhances the trade-off between tracking performance and chattering suppression. Lyapunov analysis proved the ultimate uniform boundedness of the tracking error. Experimental results demonstrated clear improvements over the baseline method, with RMSE reductions of 34.52$\%$ and 31.11$\%$, ITAE reductions of 33.79$\%$ and 32.97$\%$, and ISCT reductions of 6.69$\%$ and 17.77$\%$ for the two joints, respectively. Additional comparisons with recently reported adaptive laws confirmed faster adaptive response, smaller gain fluctuation, and improved chattering suppression. Repeated payload tests further supported the robustness and repeatability of the proposed method.

\section*{Appendix}
\vspace{-0.2ex}
\section*{Stability Analysis}

Consider the following Lyapunov function candidate:
\begin{equation}
V = \frac{1}{2} \sum_{i=1}^n s_i^2 + \frac{1}{2} \sum_{i=1}^n \tilde{\theta}_i^2.
\label{eq:v_equation}
\end{equation}
where $\tilde{\theta}_i = \theta_{i,\max} - \hat{\theta}_i \geqslant 0$. Differentiating $V$ with respect to time and substitute (\ref{eq:t_m0}) and (\ref{eq:t_m4}) yields:
\begin{align}
\dot{V} 
&= \sum_{i=1}^n \left[ -s_i \left( -\tilde{\varepsilon}_i + \tilde{\beta}_{1,i} s_i + \tilde{\beta}_{2,i} \operatorname{sig}(s_i)^{\eta_i} \right)
- \hat{K}_i |s_i| - \tilde{\theta}_i \dot{\hat{\theta}}_i \right].
\label{eq:Vdot}
\end{align}
where $\bar{\varepsilon}_i = \bar{M}_i^{-1} \left[ \left( \tau_m - \bar{M} \ddot{q} \right)_i - \left( \tau_{m(t - l_s)} - \hat{M}_i \ddot{q}_{(t - l_s)} \right)_i \right]$ is bounded according to \cite{8186211} and $\tilde{\varepsilon}_i = \bar{\varepsilon}_i - k_{3,i}\hat{\theta}_i^{\hat{\sigma}} \ddot{q}_{d,i}$, 
$\tilde{\beta}_{1,i} = \left(1 + k_{3,i}\hat{\theta}_i^{\hat{\sigma}} \right){\beta}_{1,i}$, $\tilde{\beta}_{2,i} = 
 \left(1 + k_{3,i} \hat{\theta}_i^{\hat{\sigma}}\right){\beta}_{2,i}$, $\hat{K}_i = \hat{k}_i \left(1 + k_{3,i} \hat{\theta}_i^{\hat{\sigma}} \right)$. Consider the boundary of ${\bar \varepsilon _i}$ and Assumption 1 together, we can deduce that ${\tilde \varepsilon _i}$ is bounded.

Afterwards, we have two cases to consider by substituting the proposed adaptive algorithm (\ref{eq:M}-\ref{eq:thetadot_cases}) into (\ref{eq:Vdot}). For case one: $\tilde{\theta}_i \, \dot{\hat{\theta}}_i \geqslant 0$, i.e. under the condition 1) $\hat{\theta}_i \geqslant \hat{\theta}_{i,\max}$ or $\hat{\theta}_i \leqslant 0$ and the condition 2) $0 < \hat{\theta}_i < \hat{\theta}_{i,\max}$, $|s_i| \geqslant \Omega_i$. For case two: $\tilde{\theta}_i \, \dot{\hat{\theta}}_i < 0$, i.e. under the condition $0 < \hat{\theta}_i < \hat{\theta}_{i,\max}$, $|s_i| < \Omega_i$, we get $\dot{\hat{\theta}}_i = (|s_i| + \Delta_i)^{-1}$. Then, (\ref{eq:Vdot}) becomes the inequality term and equality term in (\ref{eq:Vdot1}), respectively.
\begin{align}
\dot{V} \leqslant\ 
& - \sum_{i=1}^n s_i \left( -\tilde{\varepsilon}_i + \tilde{\beta}_{1,i} s_i + \tilde{\beta}_{2,i} \operatorname{sig}(s_i)^{\eta_i} \right)
- \sum_{i=1}^n \hat{K}_i |s_i| \notag \\
\dot{V} =\ 
& - \sum_{i=1}^n s_i \left( -\tilde{\varepsilon}_i + \tilde{\beta}_{1,i} s_i + \tilde{\beta}_{2,i} \operatorname{sig}(s_i)^{\eta_i} \right) \notag \\
& - \sum_{i=1}^n \hat{K}_i |s_i|
+ \sum_{i=1}^n \tilde{\theta}_i (|s_i| + \Delta_i)^{-1}
\label{eq:Vdot1}
\end{align}

While the equality term appears to be a more restrictive constraint than the inequality term, the $|{s_i}|$ has been bounded by ${\Omega _i}$. For further analysis, the inequality term can be reorganized as: 
\begin{align}
\dot{V} 
&\leqslant - \sum_{i=1}^n s_i \left( -\tilde{\varepsilon}_i + \tilde{\beta}_{1,i} s_i + \tilde{\beta}_{2,i} \operatorname{sig}(s_i)^{\eta_i} \right)
- \sum_{i=1}^n \hat{K}_i |s_i| \notag \\
&\leqslant - \sum_{i=1}^n |s_i| \left( \tilde{\beta}_{1,i} |s_i| + \tilde{\beta}_{2,i} |s_i|^{\eta_i} - (|\tilde{\varepsilon}_i| - \hat{K}_i) \right).
\label{eq:Vdot2}
\end{align}

If ${\hat K_i} \geqslant |{\tilde \varepsilon _i}|$ can be ensured, (\ref{eq:Vdot2}) would become $\dot{V} \leqslant - \sum_{i=1}^n |s_i| \left( \tilde{\beta}_{1,i} |s_i| + \tilde{\beta}_{2,i} |s_i|^{\eta_i} \right)$. Then $V_{(t)}$ will continuously decrease until ${s_i} = 0$ is reached. If $\hat{K}_i < |\tilde{\varepsilon}_i|$ holds, (\ref{eq:Vdot2}) has two equivalent forms: 
\begin{align}
\dot{V} 
&\leqslant - \sum_{i=1}^n |s_i| \Bigg[ \left( \tilde{\beta}_{1,i} - \left( |\tilde{\varepsilon}_i| - \hat{K}_i \right) |s_i|^{-1} \right) |s_i| 
+ \tilde{\beta}_{2,i} |s_i|^{\eta_i} \Bigg] \notag \\
&= - \sum_{i=1}^n |s_i| \left( \tilde{\beta}_{3,i} |s_i| + \tilde{\beta}_{2,i} |s_i|^{\eta_i} \right)
\label{eq:Vdot3}
\end{align}
\begin{align}
\dot{V} 
&\leqslant - \sum_{i=1}^n |s_i| \Bigg[ 
\left( \tilde{\beta}_{1,i} |s_i| + \tilde{\beta}_{2,i} 
- \left( |\tilde{\varepsilon}_i| - \hat{K}_i \right) |s_i|^{-\eta_i} \right) 
|s_i|^{\eta_i} 
\Bigg] \notag \\
&= - \sum_{i=1}^n |s_i| \left( \tilde{\beta}_{1,i} |s_i| + \tilde{\beta}_{4,i} |s_i|^{\eta_i} \right).
\label{eq:Vdot4}
\end{align}
where $\tilde{\beta}_{3,i} = \tilde{\beta}_{1,i} - \left( |\tilde{\varepsilon}_i| - \hat{K}_i \right) |s_i|^{-1}$ and $\tilde{\beta}_{4,i} = \tilde{\beta}_{2,i} - \left( |\tilde{\varepsilon}_i| - \hat{K}_i \right) |s_i|^{-\eta_i}$. When $\tilde{\beta}_{3,i} > 0$ and $\tilde{\beta}_{4,i} > 0$ hold, $V_{(t)}$ would strictly decrease as long as ${s_i} \ne 0$. Therefore, ${s_i}$ ultimately remain within the region: 
\begin{align}
|s_i| 
&\leqslant \min \left\{
\frac{|\tilde{\varepsilon}_i| - \hat{K}_i}{(1 + k_{3,i} \hat{\theta}_i^{\hat{\sigma}})\, \beta_{1,i}},
\left( \frac{|\tilde{\varepsilon}_i| - \hat{K}_i}{(1 + k_{3,i} \hat{\theta}_i^{\hat{\sigma}})\, \beta_{2,i}} \right)^{\frac{1}{\eta_i}}
\right\} = \bar{\Delta}_i.
\label{eq:si_bound}
\end{align}

\textit{Remark 5:} Conservatism and possible relaxation of the condition ${\hat K_i} \ge |{\tilde \varepsilon _i}|$. The condition may be conservative, because ${\tilde \varepsilon _i}$ aggregates the TDE mismatch and disturbance-related terms, whose instantaneous upper bound is generally difficult to determine a priori and may increase significantly under large and fast-varying disturbances. The stability result developed in this paper does not rely on this condition as a necessary requirement. Even when ${\hat K_i} < |{\tilde \varepsilon _i}|$, the closed-loop system remains ultimately uniformly bounded, and the sliding variable $s_i$ converges to the compact set characterized in \eqref{eq:si_bound}. Therefore, the conservatism of this requirement can be relaxed if an ultimately uniformly bounded convergence result is accepted, with an explicit bound determined by the difference between $|{\tilde \varepsilon _i}|$ and ${\hat K_i}$.

Then, during the sliding mode phase, \eqref{eq:s} has two equivalent representations:
\begin{align}
\dot{e}_i+ k_{1,i} D^{\chi_{1,i}} \left[ \operatorname{sig}^{\varphi_{1,i}}(e_i) \right]+\tilde{k}_{2,i} 
D^{\chi_{2,i} - 1} \left[ \operatorname{sig}^{\varphi_{2,i}}(e_i) \right] = 0
\label{eq:e_equation1}
\end{align}
\begin{align}
\dot{e}_i+ \tilde{k}_{1,i} D^{\chi_{1,i}} \left[ \operatorname{sig}^{\varphi_{1,i}}(e_i) \right]+k_{2,i} 
D^{\chi_{2,i} - 1} \left[ \operatorname{sig}^{\varphi_{2,i}}(e_i) \right] = 0.
\label{eq:e_equation2}
\end{align}
where $\tilde{k}_{1,i}=k_{1,i} - s_i \big( D^{\chi_{1,i}} [\, \operatorname{sig}^{\varphi_{1,i}}(e_i) \,] \big)^{-1}$, $\tilde{k}_{2,i}=k_{2,i} - s_i \big( D^{\chi_{2,i} - 1} \left[\, \operatorname{sig}^{\varphi_{2,i}}(e_i) \,\right] \big)^{-1}$.

For \eqref{eq:e_equation1}, when $\tilde{k}_{2,i}>0$, it will remain the FONTSM form shown in \eqref{eq:s}. In term of $|s_i| \le \bar{\Delta}_i$, the system state will reach the bounded sliding mode manifold until
\begin{align}
\left| D^{\chi_{2,i} - 1} \left[\, \operatorname{sig}^{\varphi_{2,i}}(e_i) \,\right] \right| 
\le k_{2,i}^{-1} \bar{\Delta}_i.
\label{eq:until_conditon}
\end{align}
Applying \textit{\textbf{Lemma 1}} setting choose $p = \infty $, it follows that
\begin{align}
\operatorname{ess\,sup} \big| D^{\chi_{2,i} - 1} \big[\, \operatorname{sig}^{\varphi_{2,i}}(e_i) \,\big] \big|\le 
K_i \operatorname{ess\,sup} |e_i|^{\varphi_{2,i}} 
\label{eq:essup1}
\end{align}
\begin{align}
\big| D^{\chi_{2,i} - 1} \left[\, \operatorname{sig}^{\varphi_{2,i}}(e_i) \,\right] \big| 
\le 
\operatorname{ess\,sup} \big| D^{\chi_{2,i} - 1} \left[\, \operatorname{sig}^{\varphi_{2,i}}(e_i) \,\right] \big|
\label{eq:essup2}
\end{align}
For notational simplicity, let $|e_i|_{\max}^{\varphi_{2,i}} = \operatorname{ess\,sup} |e_i|^{\varphi_{2,i}}$. It then follows from \eqref{eq:essup1} and \eqref{eq:essup2} that
\begin{align}
\big| D^{\chi_{2,i} - 1} \left[\, \operatorname{sig}^{\varphi_{2,i}}(e_i) \,\right] \big| 
\le K_i |e_i|_{\max}^{\varphi_{2,i}}.
\label{eq:essup3}
\end{align}

Let $\delta \ge 1$ be a bounded time-varying factor. Then, \eqref{eq:essup3} can be expressed as:
\begin{align}
\big| D^{\chi_{2,i} - 1} \left[\, \operatorname{sig}^{\varphi_{2,i}}(e_i) \,\right] \big|
= \delta^{-1} K_i |e_i|_{\max}^{\varphi_{2,i}}
\label{eq:essup4}
\end{align}

Substituting \eqref{eq:essup4} into \eqref{eq:until_conditon}, yields
\begin{align}
|e_i| \le |e_i|_{\max} \le 
\left( k_{2,i}^{-1} \bar{\Delta}_i \, \delta \, K_i^{-1} \right)^{1 / \varphi_{2,i}}
\label{eq:e_bound1}
\end{align}

In the same vein, the analysis of \eqref{eq:e_equation2} yields:
\begin{align}
|e_i| \le |e_i|_{\max} \le 
\left( k_{1,i}^{-1} \bar{\Delta}_i \, \delta \, K_i^{-1} \right)^{1 / \varphi_{1,i}}.
\label{eq:e_bound2}
\end{align}

Therefore, it is proven that the theoretical control errors are ultimately bounded, with the bound given by:
\begin{align}
|e_i| \le \Delta_{e_i} = 
\min \left\{
\left( \frac{ \delta\, \bar{\Delta}_i }{ k_{1,i} K_i } \right)^{\frac{1}{\varphi_{1,i}}},
\left( \frac{ \delta\, \bar{\Delta}_i }{ k_{2,i} K_i } \right)^{\frac{1}{\varphi_{2,i}}}
\right\}.
\label{eq:e_bound}
\end{align}

It can be concluded that sliding surface $s_i$ ultimately converges to the region $\bar{\Delta}_i$ and the trajectory tracking error is bounded within the region $\Delta_{e_i}$.

% References
\bibliographystyle{Bibliography/IEEEtran}
\bibliography{Bibliography/reference} %IEEEabrv instead of IEEEfull
\end{document}